\documentclass[11pt]{article}
\usepackage[a4paper,margin=1in]{geometry}
\usepackage{amsmath,amssymb,amsthm,mathtools}
\usepackage{microtype,booktabs,xcolor}
\usepackage{cite}
\usepackage{tabularx}
\usepackage{algorithm}
\usepackage[noend]{algpseudocode}
\usepackage{tikz}
\usetikzlibrary{arrows.meta,positioning,calc,decorations.pathreplacing}
\usepackage{authblk}
\usepackage[colorlinks=true,linkcolor=blue!55!black,citecolor=blue!55!black,urlcolor=blue!55!black]{hyperref}

\newtheorem{theorem}{Theorem}[section]
\newtheorem{lemma}[theorem]{Lemma}

\theoremstyle{definition}

\newcommand{\E}{\mathbb E}
\newcommand{\OPT}{\operatorname{OPT}}
\newcommand{\ALG}{\operatorname{ALG}}
\newcommand{\RoE}{\operatorname{RoE}}
\newcommand{\pos}[1]{\left(#1\right)_+}
\newcommand{\cost}{\operatorname{cost}}

\usepackage{tcolorbox}
\tcbuselibrary{skins,breakable}
\usepackage{caption}
\definecolor{PaperBlue}{HTML}{315B88}
\definecolor{PaperTeal}{HTML}{21776F}
\definecolor{PaperAmber}{HTML}{A46422}
\definecolor{PaperPurple}{HTML}{705A87}
\definecolor{PaperInk}{HTML}{243449}
\hypersetup{linkcolor=PaperBlue,citecolor=PaperTeal,urlcolor=PaperBlue}
\tcbset{paperstatement/.style={enhanced,sharp corners,
  boxrule=.45pt,left=5pt,right=5pt,top=3pt,bottom=3pt,
  before skip=8pt,after skip=8pt}}
\tcolorboxenvironment{theorem}{paperstatement,
  colback=PaperTeal!5,colframe=PaperTeal!75!black}
\tcolorboxenvironment{lemma}{paperstatement,frame hidden,
  borderline west={1.5pt}{0pt}{PaperBlue},colback=PaperBlue!3}
\tcolorboxenvironment{proposition}{paperstatement,frame hidden,
  borderline west={1.5pt}{0pt}{PaperPurple},colback=PaperPurple!4}
\tcolorboxenvironment{corollary}{paperstatement,
  colback=PaperTeal!3,colframe=PaperTeal!50}
\tcolorboxenvironment{definition}{paperstatement,frame hidden,
  borderline west={1.5pt}{0pt}{PaperAmber},colback=PaperAmber!4}
\title{Online Network Design with Additive Delays\\and Poisson Arrivals}
\author{Tianhang Lu, Runtian Ren, Shengcai Liu}
\affil{Guangdong Provincial Key Laboratory of Brain-Inspired Intelligent Computation,\\
Department of Computer Science and Engineering,\\
Southern University of Science and Technology, Shenzhen 518055, China\\
liusc3@sustech.edu.cn}
\date{}

\begin{document}
\maketitle

\begin{abstract}
We study online service systems in which requests arrive at locations
in a network. A service may open a facility and connect customers,
buy a subtree linking them to a root, or send a mobile server to visit
them. Combining requests can save opening, network, or travel costs,
but each request pays its own waiting time. The objective is the sum
of these spatial costs and individual delays.

We assume independent Poisson arrivals of known rates. The algorithms
know how often requests occur, but observe them only upon arrival.
We show that these rates suffice to make fixed spatial choices that
compete with an offline optimum free to adapt to the realized input.
We obtain long-run ratios of expectations below $27.249$ for facility
location in finite metrics, $94.415$ for multi-root aggregation on
trees, and $24.326$ for online service on trees. All three algorithms
are deterministic; the multi-root algorithm also has polynomial-time
preprocessing on rational inputs. The proofs connect rate-based
spatial choices with lower bounds on unavoidable purchase, movement,
and waiting costs. All main proofs are self-contained, and every
guarantee includes the full cost of completing finite inputs.
\end{abstract}

\section{Introduction}\label{sec:intro}
Consider a service provider receiving orders from customers across a region.
An order can be handled as soon as it arrives, but doing so may require
opening a depot, buying a connecting network, or sending a vehicle
on a long trip for a single customer. Waiting
allows nearby orders to share an opening or part of the route. The
saving comes at a price: every customer left waiting incurs a delay
cost. A good service policy must decide which orders to combine,
where to serve them, and how long to wait before committing to service.

We study three versions of this choice. In \emph{facility location with
delay}, each service pays to open a facility and to connect every
customer in its batch individually. Facilities are ephemeral: opening
one again costs again. In \emph{multi-root aggregation}, a service
chooses one root and buys a connected subtree linking its customers
to that root, sharing the cost of common edges. Roots may have
nonuniform activation costs. In \emph{online service}, a mobile server
visits customers and retains its final position between services. In all
three models, each request pays its own waiting time, and these payments
are added to the spatial cost.

The distinction between choosing service times and choosing service
locations is essential. If every client is assigned a facility or
root in advance, its temporal decisions can exploit the aggregate
demand assigned there. But the unrestricted offline optimum knows
the entire realized sequence. It may send two occurrences at the
same location to different facilities, or change the root of a batch
when its realized composition changes. Thus a fixed assignment is
an algorithmic restriction that needs to be justified against a
comparator without that restriction. For a mobile server, even choosing
a fixed home creates a restriction: returning there after every tour
may pay for movement that an unrestricted trajectory avoids.

Known demand rates offer a way to make this choice before arrivals.
We assume that each location generates an independent homogeneous
Poisson process of known rate. The rates describe demand, but reveal
no future arrival times. This stationary model isolates the question
of whether statistical information can support constant guarantees
despite the offline optimum's freedom to adapt its spatial choices.
The stochastic matching and rooted aggregation work of Mari,
Paw\l{}owski, Ren, and Sankowski~\cite{mari-et-al-2023,mari-et-al-2024}
provides the starting point: these papers report constant ratios of
expected online and offline costs under independent arrivals. Here the
additional difficulty is choosing facilities, roots, or a home for a
mobile server. We ask whether the rates can justify these fixed spatial
choices against an optimum that adapts to the realized input.

\subsection{Our contributions}
We compare expected online and offline costs over increasingly long
horizons on a fixed finite instance. The offline optimum knows all
realized arrivals and is free to choose every facility, root, or move.
All three algorithms are deterministic and complete every finite input
without being told when arrivals stop. Our main results are as follows.
\begin{itemize}
\item \textbf{Facility location in finite metrics.}
We obtain a ratio below $27.249$, allowing nonuniform facility opening
costs. The algorithm assigns each client location to a facility before
arrivals begin. The guarantee holds even though the optimum can send
different occurrences at the same location to different facilities.
Thus a fixed assignment based on rates loses only a constant factor.
\item \textbf{Multi-root aggregation on trees.}
We obtain a ratio below $94.415$, with nonuniform root activation
costs and polynomial-time preprocessing for rational inputs. The
algorithm chooses one root for each client location, whereas the
offline optimum may choose a new root for every batch. Both the root
assignment and the service schedule are selected deterministically.
\item \textbf{Online service on trees.}
We obtain a ratio below $24.326$ when at least two sites have positive
arrival rates. The server chooses a home using the rates and returns
there after each tour. The comparator starts at the same specified
location but can follow any trajectory. The result shows that this
simple fixed-home structure costs only a constant factor, despite
restricting where the online server waits between tours.
\end{itemize}

\begin{table}[t]
\centering\small
\renewcommand{\arraystretch}{1.2}
\begin{tabularx}{\textwidth}{@{}Xc@{}}
\toprule
\textbf{Model} & \textbf{Long-run ratio bound}\\
\midrule
Facility location in finite metrics & $<27.249$\\
Multi-root aggregation on trees & $<94.415$\\
Online service on trees & $<24.326$\\
\bottomrule
\end{tabularx}
\caption{Deterministic guarantees against unrestricted offline optima,
including all finite-input completion costs. Multi-root preprocessing
is polynomial in the binary input length for rational data. We do not
claim polynomial bit complexity for the other two constructions.
The mobile ratio concerns at least two positive-rate sites; empty
and single-site demand are handled directly.}
\label{tab:results}
\end{table}

All three main proofs are given in this paper. The multi-root argument
includes its own scheduling lemma for rooted forests, so no theorem
from the prescribed-resource companion is needed. The numerical
certificate for one spatial comparison is retained in the appendix.

\subsection{Our techniques}
The common idea is to use rates to decide where service should take
place, then coordinate when requests use that service. The hard part
is justifying a fixed spatial choice against an optimum that sees the
future. The three models require different ways of measuring what
that optimum must pay.

\paragraph{Facilities: balance distance against shared openings.}
Nearby demand makes an expensive facility worthwhile only if enough
requests can share its openings. We measure this tradeoff using local
demand and distance, and select facilities that represent separated
regions. The resulting assignment has a cost bound expressed through
local client weights. For the lower bound, we retain requests whose
nearby demand is not too crowded. Any offline batch must pay for these
weights through its opening, its connections, or its requests' waiting.
The same weights therefore compare both schedules.

\paragraph{Roots: choose connected regions before scheduling.}
On a tree, independent root choices can send neighboring clients in
opposite directions. We instead assign connected regions to roots,
using a specialization of the connected assignment method of
Englert et al.~\cite{connected}. An offline batch can be split among
these regions at its original time: individual waits are preserved,
and the extra spatial cost is bounded. Each region then becomes a
rooted scheduling problem. A common cost certificate lets us choose
one good assignment efficiently, before any arrivals are observed.

\paragraph{A mobile server: compare the two sides of each edge.}
A home chosen from demand rates makes the algorithm easy to schedule,
but the offline server need not return there. We choose the home so
that no branch contains more than half the demand. Across every edge,
there is then enough demand on both sides to force movement or waiting,
regardless of where the offline server chooses to stay. We combine
these edge comparisons without counting any request's delay more
than once. Coordinated tours from the home are bounded by the same
quantity, giving the constant comparison.

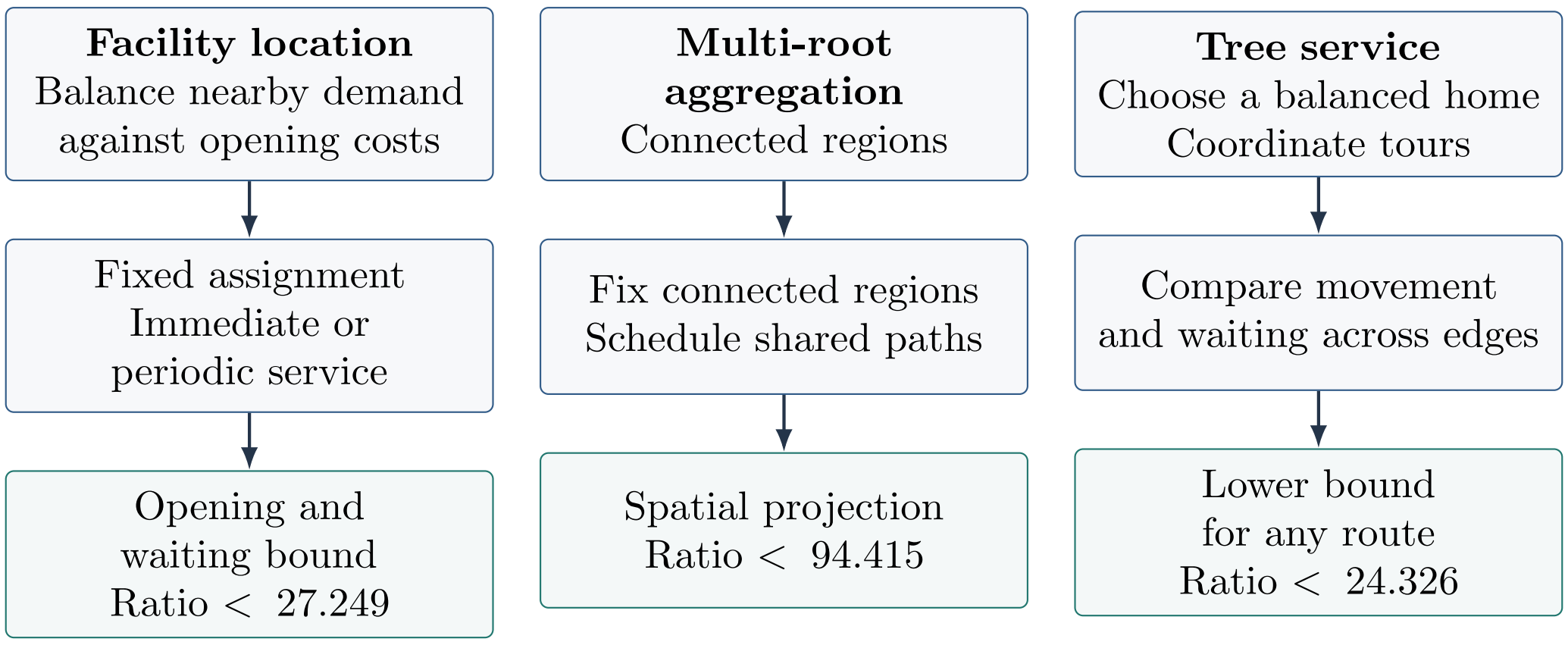
\begin{figure}[t]
\centering
\begin{tikzpicture}[
 box/.style={draw=PaperBlue,fill=PaperBlue!4,rounded corners=2pt,
 text width=3.9cm,minimum height=1.35cm,align=center,inner sep=5pt,font=\small},
 result/.style={box,draw=PaperTeal,fill=PaperTeal!5},
 link/.style={-{Latex[length=2mm]},thick,draw=PaperInk}]
\node[box] (f1) {\textbf{Facility location}\\Balance nearby demand\\against opening costs};
\node[box,right=0.4cm of f1] (r1) {\textbf{Multi-root}\\\textbf{aggregation}\\Connected regions};
\node[box,right=0.4cm of r1] (s1) {\textbf{Tree service}\\Choose a balanced home\\Coordinate tours};
\node[box,below=0.5cm of f1] (f2) {Fixed assignment\\Immediate or\\periodic service};
\node[box,below=0.5cm of r1] (r2) {Fix connected regions\\Schedule shared paths};
\node[box,below=0.5cm of s1] (s2) {Compare movement\\and waiting across edges};
\node[result,below=0.5cm of f2] (f3) {Opening and waiting bound\\Ratio $<27.249$};
\node[result,below=0.5cm of r2] (r3) {Spatial projection\\Ratio $<94.415$};
\node[result,below=0.5cm of s2] (s3) {Lower bound for any route\\Ratio $<24.326$};
\draw[link] (f1)--(f2); \draw[link] (f2)--(f3);
\draw[link] (r1)--(r2); \draw[link] (r2)--(r3);
\draw[link] (s1)--(s2); \draw[link] (s2)--(s3);
\end{tikzpicture}
\caption{Three spatial choices guided by arrival rates. Each column has a
complete local analysis; the forest scheduler and its offline
certificate are proved in Section~\ref{sec:forest-scheduling}. The
multi-root guarantee includes deterministic polynomial preprocessing.}
\label{fig:spatial-framework}
\end{figure}

\subsection{Related work}
\paragraph{Aggregation and delay frameworks.}
Multi-level aggregation includes acknowledgment and joint
replenishment on shallow trees. Bienkowski et al.~\cite{bienkowski-et-al-2016}
study online multi-level aggregation, and Buchbinder
et al.~\cite{buchbinder-2017} give depth-dependent guarantees for
the deadline variant. Azar and Touitou's frameworks
\cite{azar-touitou-2019,azar-touitou-2020} connect aggregation,
facility location, and other network design problems with delay or
deadlines to common online constructions and offline primitives.
These results concern adversarial inputs. Our statistical assumption
and ratio-of-expectations criterion are different, and our constant
bounds do not imply constant adversarial guarantees.

\paragraph{Facility location.}
For linear delay, Bienkowski et al.~\cite{bienkowski-et-al-2022}
obtain an $O(\log m/\log\log m)$ guarantee in their one-sided model
and a constant guarantee in a two-sided model with different
connection and delay charges. We use ephemeral openings and one
waiting cost per request. Our facility theorem allows arbitrary
finite metrics and nonuniform opening costs; it compares a fixed
assignment with an offline optimum free to change facilities between
occurrences. The radius and space--time arguments are given in full.

\paragraph{Mobile service.}
Online service with delay was introduced by Azar, Ganesh, Ge, and
Panigrahi~\cite{azar-ganesh-ge-panigrahi-2017}. Azar and
Touitou~\cite{azar-touitou-2019} developed a framework for metric
optimization with delay or deadlines and obtained a randomized
$O(\log^2 n)$ guarantee for service in general metrics. Touitou's
subsequent deterministic algorithm gives $O(\log n)$~\cite{touitou-2023}.
These guarantees apply to adversarial arrivals and address a broader
metric domain. Our constant bound uses known Poisson rates and the cut
structure of trees; it does not give a constant guarantee for adversarial
arrivals on arbitrary trees or for stochastic service in general metrics.

Information about arrival rates is also distinct from information about
future delay functions. Touitou~\cite{touitou-service-2025} studies
nonclairvoyant service with delay. Here the delay function of each request
is the known linear function; only the future arrival sequence is unknown.
The algorithm's additional information consists of stationary rates.

\paragraph{Stochastic arrivals and spatial projection.}
Mari et al.~\cite{mari-et-al-2023,mari-et-al-2024} study stochastic
matching and rooted aggregation with independent arrival streams.
Their aggregation work combines immediate service for infrequent
demand with periodic service for frequent demand and motivates
extensions to facility location. The connected $0$-extension
construction we specialize is due to Englert et al.~\cite{connected}.
% Our contribution in this use of their method is the finite tree
% implementation, explicit $313/8$ service bound, and polynomial
% local-marginal construction.
% The companion work of Lu, Ren, and Liu~\cite{resource-companion}
% studies arbitrary prescribed resource sets and sharper scalar guarantees.
The path-budget LP and synchronized-clock techniques used here are
closely related. For the rooted forests needed in this paper, we give
the full scheduling proof with an elementary scalar estimate; none of
our main theorems invokes a result from that companion.

\subsection{Organization}
Section~\ref{sec:model} gives the models and basic probability facts.
Section~\ref{sec:spatial-story} explains all three algorithms, their
lower-bound ideas, and how the constants follow. These sections give
the complete main story before the technical development.
Section~\ref{sec:technical-tools} records the probability tools.
Sections~\ref{sec:facility}, \ref{ag:sec:multi-model}, and
\ref{sec:mobile} give the full facility, multi-root, and tree-service
proofs, respectively. Appendix~\ref{app:partition} contains the
connected-assignment analysis and its numerical certificate.

\section{Models and basic Poisson facts}\label{sec:model}
\subsection{The three service models}
The spatial instance and set of request types are finite. A type
specifies a client location; distinct arrivals at that location are
distinct requests. Request $i$ arrives at $a_i$ and is served at
$t_i\ge a_i$, incurring waiting cost $t_i-a_i$. In every model we
minimize spatial cost plus the sum of these individual waiting costs.
We write $d(u,v)$ for distance and $w(S)$ for the sum of edge costs
in an edge set $S$.

\paragraph{Facility location.}
Possible clients and facilities lie in a finite metric. Facility $f$
has opening cost $c_f\ge0$. A service chooses a facility and a batch
of pending requests, pays the opening cost once, and pays the distance
to the facility separately for every request. The facility closes
after the service. A later service pays another opening cost.

\paragraph{Multi-root aggregation.}
Clients lie at vertices of a finite tree with nonnegative edge costs.
There is a nonempty set of available roots, with activation cost
$c_r\ge0$ for root $r$. A service chooses a root and a connected
subtree containing that root and its served clients. It pays the
activation once and each edge of the subtree once. Later services
purchase their networks again. Different batches may choose different
roots, including in the offline optimum.

\paragraph{Online service.}
A single server starts at a specified vertex $s_0$ of a finite tree.
Traversing an edge costs its length, and visiting a vertex serves all
pending requests there for free. The server retains its position
between services and may move proactively. The offline server starts
at the same $s_0$ and has no obligation to return to any home.
Zero-length edges may be contracted.

All services and moves take zero time; there are no capacities.
Arrivals at a service time are observed first, and simultaneous
actions are processed in execution order. Splitting a batch into
simultaneous services preserves additive waiting, though its spatial
cost may change. This distinction will be useful for root assignments.

\subsection{Arrivals and performance}
Type $x$ follows a homogeneous Poisson process of known rate
$\lambda_x\ge0$, independently of other types. Equivalently, its
successive interarrival times are independent exponentials of mean
$1/\lambda_x$ when the rate is positive. The algorithm knows the rates
and spatial instance, but sees requests only upon arrival. Zero-rate
types can be discarded; their locations remain available for spatial
constructions and travel.

Let $I_T$ contain the requests arriving in $[0,T]$. The algorithm
receives no end-of-input signal and must eventually complete all
requests. Its cost $\ALG_T=\ALG(I_T)$ includes all service, movement,
and waiting needed after $T$. The offline optimum $\OPT_T$ completes
the same input with full knowledge of the realized arrivals. For a
fixed instance with positive optimal asymptotic cost, we use
\begin{equation}\label{eq:roe}
 \RoE(\ALG)=\limsup_{T\to\infty}
 \frac{\E[\ALG_T]}{\E[\OPT_T]}.
\end{equation}
This compares expected total costs, not expected samplewise ratios.
We write $O_I(1)$ for a finite constant depending on the instance
but not on $T$. Our estimates include completion before taking limits.

\subsection{Basic Poisson calculations}
A fixed interval of length $p$ contains $\lambda p$ arrivals in
expectation and is empty with probability $e^{-\lambda p}$. Counts
in disjoint intervals are independent. Combining independent streams
adds their rates; independently retaining each arrival with probability
$u$ gives a Poisson stream of rate $u\lambda$.

Conditional on a fixed number of arrivals in an interval, their
unordered times are independent uniform points. Hence collecting
rate-$\lambda$ arrivals for a fixed period $p$ and serving at its end
incurs expected waiting $\lambda p^2/2$, or rate $\lambda p/2$ over
repeated periods. This is the elementary waiting calculation used
below. These facts concern intervals fixed independently of arrivals;
we do not assume they hold for batches chosen after seeing the future.
The more general identities needed for lower bounds are proved in
Section~\ref{sec:technical-tools}.

\section{How the three guarantees are obtained}\label{sec:spatial-story}
The algorithms make their main spatial choices before requests arrive:
an assignment to facilities, an assignment to roots, or a home for the
server. This section explains why those restrictions are affordable.
In each case we identify a quantity that bounds both the algorithm's
expected cost and the unrestricted optimum's expected cost. The
technical sections establish the exact constructions and estimates.

\subsection{Facilities: choose where nearby demand should meet}
Two nearby clients may benefit from sharing an expensive facility
opening. Assigning each to its nearest facility can miss this saving;
assigning everyone together can instead create excessive connection
cost. Arrival rates help decide how much nearby demand is available
to share an opening.

\paragraph{The spatial choice.}
Around each facility we grow a radius until nearby demand is sufficient
to balance its opening cost. Expensive facilities and sparse regions
need larger radii. We select representatives in increasing order of
these scales, suppressing facilities too close to one already selected.
Each client is assigned to a representative that balances distance
against the cost of supporting its openings. This uses only the rates,
distances, and opening costs, not future requests.

The construction gives each client $x$ a weight $\alpha_x$ measuring
its local service scale. A nearby cheap facility makes this weight
small; expensive service or sparse demand makes it larger. Let
$A=\sum_x\lambda_x\alpha_x$. The weights are chosen jointly, so they
account for opportunities to share openings rather than charging
every client for an independent facility.

\paragraph{When to serve.}
After fixing the assignment, the requests assigned to a facility
form one Poisson stream. Connections cost the same regardless of
when those requests are served. Only openings and waiting need to
be balanced. If the assigned rate is $\Lambda$ and the opening cost
is $c$, immediate service has opening cost rate $c\Lambda$.
Service every $p$ time units has opening-plus-waiting rate at most
\[
 \frac cp+\frac{\Lambda p}{2}.
\]
The algorithm chooses the cheaper mode and skips empty batches. The
radius construction bounds its connections, openings, and waiting
together by $19A/4$ per unit time, apart from a finite completion cost.

\paragraph{Why the offline optimum must also pay.}
It would be incorrect simply to charge $\alpha_x$ for every arrival:
a single offline opening could serve many nearby requests cheaply.
The analysis therefore retains only requests whose neighborhoods in
space and time are not too crowded. These retained requests carry a
constant fraction of the total weight in expectation, by the Poisson
arrival law. This selection is used only in the proof.

Now fix any offline batch. A retained request that is far from its
facility, or waits a long time, pays for its weight through its own
connection or waiting. The remaining retained requests are nearby
and recently arrived. The crowding condition limits their combined
weight, so the batch's single opening pays for them. This is a
comparison with every realized schedule, including one chosen with
full knowledge of the input.

The resulting pair of bounds is
\[
 \E[\ALG_T]\le\frac{19}{4}AT+O_I(1),
 \qquad
 \E[\OPT_T]\ge\frac{25}{32}e^{-3/2}AT.
\]
Dividing gives the ratio below $27.249$. The facility count and the
variation in opening costs enter through the same $A$, and hence do
not create extra factors. Theorem~\ref{fl:thm:main} proves the
guarantee. The proof includes the radii, the assignment rule, the
retention estimate, and the last clearing after arrivals stop.

\subsection{Multiple roots: keep each assigned region connected}
A tree service may choose a root afresh for every batch. A fixed
assignment could be much more restrictive if nearby clients were
sent toward different roots along overlapping paths. Connected
regions avoid this problem: each client uses the path to the root
of its own region, and different regions have disjoint edges.

\paragraph{Comparing spatial choices.}
We first represent each root activation cost by an extra edge ending
at that root's terminal. Purchasing a path to the terminal then pays
the activation exactly once. We specialize the connected assignment
construction of Englert et al.~\cite{connected} to partition this
augmented tree into regions, one per terminal.

For any fixed offline service, split its requests by assigned root
and serve the resulting batches at the original time. Every request
keeps exactly the same waiting cost. The connected assignment bounds
the increase in spatial cost, averaged over a finite distribution of
partitions, by $313/8$. The same distribution works for every input
and is chosen before arrivals. This step is where additive delay is
essential: splitting a batch does not repeat its requests' waits.

\paragraph{Scheduling one chosen forest.}
Once the regions are fixed, every client has a prescribed root path.
We schedule these paths using a rooted-forest lemma proved in this
paper. Its LP distributes edge purchase costs and request waiting
costs among simpler time scales, without charging either budget more
than once. A direct one-resource argument supplies the lower bound:
between consecutive arrivals, the schedule must either buy another
service or pay enough waiting. Poisson gap probabilities turn this
observation into an expected cost certificate.

The LP dual supplies service periods. We align them so that a longer
period is an integer multiple of each shorter one. Requests due
together share their root edges; empty queues buy nothing. Both the
offline lower bound and the online upper bound are expressed through
the same LP value. Their ratio gives a rooted-forest scheduling
factor $1431/593$. No result from the prescribed-resource companion
is invoked for this step.

\paragraph{Selecting one assignment efficiently.}
The spatial comparison initially averages over partitions, whereas
our final algorithm must choose one. We do not search through all
connected assignments. On a tree, compatible labels at vertices and
edges describe connected root regions. A polynomial-size LP and a
decomposition of its solution produce a polynomial-size list of
assignments with the required averaged spatial bound.

Each candidate forest has a computable scheduling certificate. Choose
the candidate with the smallest such value. Its value is no larger
than the average, and its online cost is controlled by that same
value. Thus we obtain one deterministic assignment from the known
instance, rather than choosing a partition after seeing arrivals.
Rational periods can also be selected in polynomial time.

The two losses have separate meanings: one pays for fixing the roots,
and the other pays for online scheduling within those fixed regions.
Together they give
\[
 \frac{313}{8}\cdot\frac{1431}{593}<94.415.
\]
Theorem~\ref{thm:efficient-roots} gives the full deterministic
polynomial guarantee. Section~\ref{sec:forest-scheduling} proves
the temporal lemma, Section~\ref{sec:polynomial-roots} proves the
efficient selection, and Appendix~\ref{app:partition} proves the
spatial comparison, including its finite arithmetic certificate.

\subsection{A mobile server: choose a balanced home}
A mobile server carries its position from one service to the next.
Repeatedly returning to a poorly chosen home can waste movement that
the offline server avoids. Choosing the highest-rate vertex is not
enough: a distant group of individually less frequent clients may
have much greater combined demand.

We choose a rate-weighted centroid $h$ of the tree: every component
remaining after its removal has at most half the total rate. The
algorithm moves to $h$ initially and returns there after each tour.
The offline comparator starts at the original specified position and
retains complete freedom to move and wait elsewhere.

\paragraph{Coordinating the tours.}
Starting from the home, we assign a service scale to each vertex by
balancing the movement needed to reach it against the demand below it.
Sparse demand favors serving without much extra waiting; frequent
demand supports collecting a larger batch. The scales increase along
paths away from home. Rounding them to aligned periods means that a
tour visiting a distant due request can also serve due requests along
its path.

At a tick, the server takes the union of paths to due nonempty queues,
traverses that subtree, and returns home. Each shared edge is used
at most once in each direction. Other pending requests encountered
on the tour are served as well. Every request is served by its next
own tick, so the policy also completes a finite input without an
end signal.

\paragraph{A lower bound that does not force OPT to return home.}
Remove an edge and consider its two sides. Serving demand on both
sides eventually requires a crossing or waiting. The centroid choice
ensures that demand in the branch away from home can be matched with
enough demand outside it. The analysis extracts balanced two-site
Poisson instances from these streams and projects the original
trajectory onto the two sides of each edge.

One request can affect several edge comparisons, so adding their full
delays would overcharge OPT. We allocate only fractions of its waiting
to these comparisons and show that the total allocation is at most
its actual waiting. Crossing costs add along physical edges. The
projection therefore gives a valid lower bound even for a trajectory
that never returns home. The random selections in this argument are
analysis tools; the actual online algorithm is deterministic.

Let $B$ denote the sum of client rates weighted by their service
scales. The tour analysis and the edge projections give, respectively,
\[
 \E[\ALG_T]\le C_0BT+O_I(1),
 \qquad \E[\OPT_T]\ge\frac1{16}BT,
 \qquad C_0=1+\frac1{4(\ln2)^2}.
\]
Thus the ratio is at most $16C_0<24.326$. The finite completion
term includes the initial move from $s_0$ to $h$.
Theorem~\ref{os:thm:tree} gives the full result. Its proof first
establishes the two-site comparison, then bounds the tours, and
finally combines the edge projections without repeated charging.

\paragraph{What is common, and what is not.}
In all three models, rates justify a fixed spatial policy by connecting
its cost to a lower bound on unrestricted OPT. The lower bounds differ
because an opening, a rooted purchase, and a persistent trajectory
create different opportunities for sharing. We retain complete proofs
of all three comparisons. Polynomial bit complexity is established
for multi-root preprocessing; the facility and mobile constructions
retain their stated finite, exact-real descriptions.

\section{Probability tools for the proofs}\label{sec:technical-tools}
The preceding sections described the algorithms and the comparisons
behind their guarantees. We now give the complete technical arguments.
This section collects the Poisson identities used to average the
lower-bound constructions; subsequent sections prove the model-specific
estimates. The insertion identity counts the contribution of a request
whose weight depends on neighboring arrivals. It is an analysis tool,
not information supplied to the online algorithms.

We write $z_+=\max\{z,0\}$, use $\mathbf1_A$ for the indicator of an
event, and let $|J|$ denote the length of a measurable time set.
Charging intervals are half-open. Nonnegative terms permit exchanges
of sums, integrals, and expectations by Tonelli's theorem.

\subsection{Poisson identities used in the proofs}\label{sec:poisson}
For a rate-$\lambda$ process, the number $N(J)$ of arrivals in an
interval $J$ has distribution
$$
 \Pr(N(J)=k)=e^{-\lambda|J|}\frac{(\lambda|J|)^k}{k!},
 \qquad \E[N(J)]=\operatorname{Var}(N(J))=\lambda|J|.
$$
Counts in disjoint intervals are independent. Conditional on $n$
arrivals in a finite interval, their unordered times are $n$
independent uniform points in that interval. In particular, for a
nonnegative deterministic function $g$,
\begin{equation}\label{eq:poisson-sum}
 \E\left[\sum_{a_i\in[0,T]}g(a_i)\right]
   =\lambda\int_0^T g(t)\,dt.
\end{equation}
Conditioning on the count proves this identity: the conditional
expectation is $nT^{-1}\int_0^Tg(t)\,dt$, and $\E[n]=\lambda T$.
We use it, for example, with $g(t)$ equal to the wait until the next
deterministic clock tick. Nonnegative integrands also allow us to
exchange sums, integrals, and expectations by Tonelli's theorem.

For independent streams, superposition adds their rates. Conversely,
independently assigning an arrival a mark $j$ with probability $q_j$
produces independent Poisson streams of rates $q_j\lambda$. The
generating function
$\E[\prod_j z_j^{N_j}]=\exp(\lambda|J|\sum_jq_j(z_j-1))$
factors into the generating functions of these independent counts.
These statements apply to marks chosen independently for each
occurrence; arbitrary routing based on the future need not preserve
a Poisson law.

The void probability is $\Pr(N(J)=0)=e^{-\lambda|J|}$.
It remains valid for a deterministic interval that depends on the
instance parameters. We will not apply it to intervals selected by
an offline schedule after observing arrivals. Instead, our lower
bounds first establish an inequality for every realized schedule
and then average a certificate whose counting regions are fixed
independently of that schedule.

We also use the following extension of~\eqref{eq:poisson-sum}, in
which the contribution of a distinguished arrival depends on the
other arrivals. It is useful for the local facility certificates.

\begin{lemma}[Poisson insertion]\label{fl:lem:insertion}
Fix $x$, and let $u_x(a,J)$ be a nonnegative measurable quantity defined
for a distinguished type-$x$ request at time $a$ and a finite multiset $J$
of other requests. Then
\begin{equation}\label{fl:eq:insertion}
 \E\!\left[\sum_{i\in I_T:\,x_i=x}
   u_x(a_i,I_T\setminus\{i\})\right]
 =\lambda_x\int_0^T\,\E[u_x(a,I_T)]\,da.
\end{equation}
On the right, $I_T$ is an independent Poisson input to which the
distinguished request can be added inside the definition of $u_x$.
\end{lemma}
\begin{proof}
Condition on the number $N_x=n$ of type-$x$ arrivals. Their times are
independent uniforms in $[0,T]$, and exchangeability makes the sum on
the left equal in expectation to $n$ times the contribution of one
distinguished uniform time, with $n-1$ other type-$x$ times. The identity
$n\Pr(N_x=n)=\lambda_xT\Pr(N_x=n-1)$ changes the latter $n-1$ points
back into an unconditional type-$x$ Poisson process. All other types
retain their independent laws. Integrating the distinguished uniform
time proves~\eqref{fl:eq:insertion}. No conditioning on an arrival at one
prescribed time is needed.
\end{proof}

\section{Proof of the facility guarantee}\label{sec:facility}
A facility service combines a shared opening with an individual
connection for every customer. It may be beneficial to send several
nearby, infrequent client streams to one facility and share its
openings; doing so also increases some connection distances. This
couples the spatial assignment to the service frequency. The
offline optimum can make that tradeoff anew for every realized
batch, whereas we will choose one assignment using only the rates.

\paragraph{Idea and illustration.}
Consider two client locations close to the same expensive facility.
Analyzing them separately ignores the saving from their combined
arrival rate. Assigning every client to its nearest facility can
also miss a worthwhile shared opening. We therefore grow a radius
according to nearby demand and the cost of an opening, rather
than choosing an assignment from distances alone. Separated
representatives give the algorithm's facilities. A related radius
at each client supplies a weight for the lower-bound proof.

The comparison with the optimum must account for congestion:
arbitrarily many requests may be close to the same offline
opening, so one opening cannot pay an unrestricted sum of their
weights. We retain a subset with bounded local crowding and show
that openings, connections, or individual waiting pay those
weights. The same potential bounds the cost of our assignment.
After the model, we construct the radii and assignment and analyze
the online service periods, including completion. We then prove the
offline certificate and combine the two estimates.

\subsection{The facility model}\label{fl:sec:preliminaries}

\paragraph{Notation.}
Let $(X\cup F,d)$ be a finite metric space, where $X$ contains possible
request locations and $F\ne\varnothing$ contains possible facility
locations. The sets may overlap. All distances are finite, and each
facility $f$ has a finite opening cost $c_f\geq0$. Nothing requires
opening costs to be uniform. Distinct requests at the same point remain
distinct occurrences. We write $\pos z=\max\{z,0\}$.

At time $t$, an algorithm may choose $f\in F$ and a nonempty subset $B$ of
pending requests whose arrival times $a_i$ satisfy $a_i\le t$. It pays
$c_f$ and connects each $i\in B$ to $f$, paying $d(x_i,f)$ for that
request. The facility closes after this event; opening it at another time
costs again. Several service events at one time are allowed, but each has
its own opening and delay charge. There is no capacity constraint or travel
time. \begin{equation}\label{fl:eq:add-cost}
\cost(B,f,t)=c_f+\sum_{i\in B}\bigl(d(x_i,f)+t-a_i\bigr).
\end{equation}
The total cost is the sum over service events. The offline optimum
may choose different facilities for different occurrences of the same type.
The algorithm knows the metric, all opening costs, and all arrival rates.

\subsection{Radii, assignment, and service periods}
\label{sec:r5-facility-dependencies}

Assume positive total arrival rate; otherwise the empty schedule suffices.
For $c_f>0$, let $R_f$ be the unique solution of
$$
 \sum_x\lambda_x(R_f-d(x,f))_+^2=c_f.
$$
The left side is continuous, strictly increasing once positive, and
unbounded. Set $R_f=0$ if $c_f=0$, and define
$$
 r_f=\min\{c_f,R_f\},\qquad
 \alpha_x=\min_f\max\{d(x,f),r_f\}.
$$
Put $A=\sum_x\lambda_x\alpha_x$. The function $\alpha$ is 1-Lipschitz.
Also $\alpha_x\le d(x,f)+c_f$ and
$(\alpha_x-d(x,f))_+\le(R_f-d(x,f))_+$ for every $f$.
Thus all three hypotheses of the facility certificate hold.

Process facilities in increasing $r_f$, breaking ties deterministically.
Select $f$ if every earlier selected facility is farther than $2r_f$.
If $g$ is rejected, let $h(g)$ be an earlier selected facility within
$2r_g$; selected facilities represent themselves.
For each $x$, choose a witness $g_x$ attaining $\alpha_x$ and assign
$\pi(x)=h(g_x)$. Then
$$
 r_{\pi(x)}\le\alpha_x,\qquad d(x,\pi(x))\le3\alpha_x.
$$
The open balls $B(f,r_f)$ of selected facilities are disjoint, since
the later facility $f'$ has distance greater than
$2r_{f'}\ge r_f+r_{f'}$ from the earlier one.

If $f$ is selected and $d=d(x,f)<r_f$, then
$$
 \alpha_x\ge\frac{2r_f-d}{3}\ge\frac23(r_f-d).
$$
Indeed a smaller potential would give $r_{g_x}<r_f$, so $h(g_x)$
was already selected when $f$ was processed, while
$d(f,h(g_x))\le d+3\alpha_x<2r_f$, a contradiction.

Temporarily serve the clients assigned to a selected facility instantly
when $r_f=c_f$, and otherwise at period $2r_f$.
Connection rate is at most $3A$.
For a periodic facility $r_f=R_f<c_f$, its opening rate is at most
$$
 \frac{c_f}{2r_f}
 =\sum_x\lambda_x\frac{(r_f-d(x,f))_+^2}{2r_f}
 \le\frac34\sum_{x\in B(f,r_f)}\lambda_x\alpha_x.
$$
Disjointness makes the total periodic opening rate at most $3A/4$.
Periodic waiting rate is at most
$\sum_{x\ \mathrm{periodic}}\lambda_x\alpha_x$.
Instant opening rate is at most
$\sum_{x\ \mathrm{instant}}\lambda_x\alpha_x$.
These last two groups are disjoint, so the total rate is at most $19A/4$.

For this fixed assignment, put $\Lambda_f=\sum_{\pi(x)=f}\lambda_x$ and
$$
 q_c(\Lambda)=\min\{c\Lambda,\sqrt{2c\Lambda}\},\qquad
 \Psi_q(\pi)=\sum_x\lambda_xd(x,\pi(x))+\sum_f q_{c_f}(\Lambda_f).
$$
The algorithm serves immediately when $c_f\Lambda_f\le2$ and
otherwise uses period $\tau_f=\sqrt{2c_f/\Lambda_f}$, skipping empty
ticks. Unused facilities are ignored. This minimizes
$\min\{c_f\Lambda_f,c_f/\tau+\Lambda_f\tau/2\}$, so the radius-period
comparison above gives $\Psi_q(\pi)\le19A/4$.
The algorithm uses the constructed assignment only; it does not
optimize over all client-to-facility assignments.

For a periodic facility, on input $I_T$, expected opening cost is at
most $c_f(T/\tau_f+1)$, including the final clearing after $T$.
Expected waiting is
$$
 \Lambda_f\int_0^T(\tau_f\lceil t/\tau_f\rceil-t)\,dt
 \le\frac{\Lambda_f\tau_f T}{2}+\frac{\Lambda_f\tau_f^2}{8}.
$$
Since $\Lambda_f\tau_f^2=2c_f$, the excess over the linear rate is
at most $5c_f/4$, and therefore at most $2c_f$.
All requests clear at their next scheduled tick without an end signal.
Thus
$$
 \mathbb E[\operatorname{ALG}(I_T)]
 \le T\Psi_q(\pi)+2\sum_{f\ \mathrm{periodic}}c_f.
$$
If $A=0$, every positive-rate type has a witness at distance zero
with radius zero. A positive-cost facility has positive radius,
so this witness is free. The assignment consequently offers
zero-cost immediate service.

\subsection{Variable-radius space--time certificates for facility location}

Suppose the nonnegative, instance-dependent vector $\alpha$ satisfies
$$
 \alpha_x\le\min_{f\in F}\{c_f+d(x,f)\},\qquad
 |\alpha_x-\alpha_y|\le d(x,y),\qquad
 \sum_y\lambda_y(\alpha_y-d(y,f))_+^2\le c_f\quad(f\in F).
$$
Write $A=\sum_x\lambda_x\alpha_x$.

\begin{theorem}[Finite-horizon facility certificate]
For every $T>0$, against the unrestricted offline optimum,
$$
 \mathbb E[\operatorname{OPT}(I_T)]
 \ge \frac{25}{32}e^{-3/2}AT.
$$
\end{theorem}
\begin{proof}
Fix
$$
 k=\frac{134}{25},\qquad \theta=\frac1k=\frac{25}{134}.
$$
For each $x$ with $\alpha_x>0$, put
$$
 m_x=\min\{c_f:d(x,f)\le\theta\alpha_x\},
$$
where the minimum of an empty set is $+\infty$.
The singleton inequality gives
$m_x\ge(1-\theta)\alpha_x>0$ whenever $m_x$ is finite.
For an occurrence $(x,a)$, define its deterministic, variable-radius region
$$
 D_x(a)=\{(y,b):d(x,y)+|b-a|\le\theta(\alpha_x+\alpha_y)\}.
$$
Let $N_x(a)$ count all requests of $I_T$ in this region, including the
occurrence itself. Retain $(x,a)$ if $m_x=+\infty$ or
$N_x(a)\le k m_x/\alpha_x$. Requests with zero potential may be discarded.
Let $W_T$ be the sum of the potentials of retained requests.

We first prove, pathwise, that $W_T\le k\operatorname{OPT}(I_T)$.
Consider any batch at facility $f$, at any time $t$, including $t>T$.
Write $d_i=d(x_i,f)$, $w_i=t-a_i$, and $\alpha_i=\alpha_{x_i}$.
Call a retained request expensive if $d_i+w_i\ge\theta\alpha_i$.
Its potential is at most $k(d_i+w_i)$.
If the batch contains cheap retained requests, choose one with maximum
potential as a pivot $p$. Every cheap retained request $i$ satisfies
$$
 d(x_i,x_p)+|a_i-a_p|
 \le d_i+d_p+|w_i-w_p|
 \le d_i+w_i+d_p+w_p
 <\theta(\alpha_i+\alpha_p),
$$
so it belongs to $D_{x_p}(a_p)$.
Also $d_p<\theta\alpha_p$, hence $m_p\le c_f<\infty$.
By maximality of $\alpha_p$ and the retention condition, the total
potential of the cheap retained requests in this batch is at most
$$
 \alpha_pN_{x_p}(a_p)\le k m_p\le k c_f.
$$
Together the cheap and expensive charges are bounded by $k$ times
the cost of the batch. Summing over any completing schedule, and then
taking the infimum over schedules, proves the asserted pathwise bound.

We next bound the retention probability of a request inserted at a
fixed $(x,a)$ into an independent Poisson input on $[0,T]$.
When $m_x=+\infty$, retention is automatic.
Otherwise choose $g\in F$ attaining $m_x$ and write $\alpha=\alpha_x$.
For a location $y$ whose time section in $D_x(a)$ is nonempty, put
$$
 u=\frac{d(x,y)}\alpha,\qquad z=\frac{\alpha_y}\alpha,\qquad
 L=\theta(1+z)-u\ge0,\qquad v=z-u-\theta,
 \qquad \eta=1-4\theta-\theta^2>0.
$$
The Lipschitz constraint implies $z+u\ge1$, and consequently
$$
 (1+\theta)v-\eta-2L=(1-\theta)(z+u-1)\ge0.
$$
In particular, $v>0$. The triangle inequality also gives
$\alpha_y-d(y,g)\ge\alpha v$.
For every $v>0$, completing a square, or maximizing a one-variable
quadratic expression, gives
$$
 \frac{2L}{v^2}
 \le\frac{(1+\theta)v-\eta}{v^2}
 \le\frac{(1+\theta)^2}{4\eta}
 =:K=\frac{(k+1)^2}{4(k^2-4k-1)}.
$$
The time section of $D_x(a)$ at $y$, after intersection with $[0,T]$,
has length at most $2\alpha L$. Thus the number $M=N_x(a)-1$ of
other requests in the region is Poisson with mean
$$
 \mu\le\sum_y2\lambda_y\alpha L_+
 \le\frac K\alpha\sum_y\lambda_y(\alpha_y-d(y,g))_+^2
 \le K\frac{m_x}{\alpha}.
$$
Only locations with $L\ge0$ need to be included in the middle bound.
The actual mean only decreases near either end of the input horizon.

Put $r=m_x/\alpha\ge1-1/k$ and $h=\lfloor kr\rfloor\ge4$.
Failure of retention means exactly $M\ge h$.
The chosen parameter satisfies
$$
 \frac Kk=\frac{632025}{2107016}<\frac3{10}=:\rho,
 \qquad
 \mu<\rho(h+1).
$$
We prove a uniform bound for all the possible integers $h$, rather
than truncating the range of $m_x/\alpha$.
Define
$$
 a_h=e^{-\rho(h+1)}\frac{[\rho(h+1)]^h}{h!},\qquad
 b_h=\frac{a_h}{1-\rho}.
$$
For a Poisson variable of mean $\rho(h+1)$, the ratio of successive
probability terms at and beyond its $h$th term is at most $\rho$.
Its upper tail at $h$ is therefore at most $b_h$.
Moreover,
$$
 \frac{b_{h+1}}{b_h}
 =\rho e^{-\rho}\left(1+\frac1{h+1}\right)^{h+1}
 <\rho e^{1-\rho}<1-\rho.
$$
For the last inequality, $e^u<(1-u)^{-1}$ for $0<u<1$ yields
$$
 e^{7/10}<\left(\frac{30}{23}\right)^3<\frac73,
$$
so $(3/10)e^{7/10}<7/10$.
It follows that for every $h\ge5$,
$$
 \Pr\{\operatorname{Pois}(\rho(h+1))\ge h\}
 \le b_h\le b_5<(1-\rho)b_4=a_4
 \le\Pr\{\operatorname{Pois}(3/2)\ge4\}.
$$
The same upper bound is tautological for $h=4$.
Monotonicity of Poisson counts in their mean therefore shows that
every inserted positive-potential request is retained with probability
at least
$$
 p=\Pr\{\operatorname{Pois}(3/2)\le3\}
 =e^{-3/2}\left(1+\frac32+\frac{(3/2)^2}{2}
                         +\frac{(3/2)^3}{6}\right)
 =\frac{67}{16}e^{-3/2}.
$$
Writing $\operatorname{ret}_x(a;I_T\cup\{(x,a)\})$ for the inserted
request's retention indicator, the Poisson insertion identity gives
$$
 \mathbb E[W_T]
 =\sum_x\lambda_x\alpha_x\int_0^T
     \mathbb E[\operatorname{ret}_x(a;I_T\cup\{(x,a)\})]\,da
 \ge pTA.
$$
Combining this with the pathwise batch bound
proves
$$
 \mathbb E[\operatorname{OPT}(I_T)]\ge\frac pk TA
 =\frac{25}{32}e^{-3/2}TA.
$$
\end{proof}

\begin{theorem}[Facility-location guarantee]\label{fl:thm:main}
For the feasible potential vector and assignment $\pi$ constructed above, the online policy satisfies
$$
 \Psi_q(\pi)
 \le \frac{152}{25}e^{3/2}
        \frac{\mathbb E[\operatorname{OPT}(I_T)]}{T},
$$
and
$$
 \mathbb E[\operatorname{ALG}(I_T)]
 \le \frac{152}{25}e^{3/2}\,\mathbb E[\operatorname{OPT}(I_T)]
       +2\sum_{f\ \mathrm{periodic}}c_f.
$$
Its long-run ratio of expectations is at most
$(152/25)e^{3/2}<27.249$.
\end{theorem}
\begin{proof}
Use the feasible-potential construction, the assignment estimate
$\Psi_q(\pi)\le19A/4$, and the finite-horizon policy estimate
$\mathbb E[\operatorname{ALG}(I_T)]\le T\Psi_q(\pi)
+2\sum_{f\ \mathrm{periodic}}c_f$.
Combining them with the theorem gives the displayed inequalities.
If $A>0$, the offline lower bound is linear in $T$, so the completion
term vanishes in the ratio as $T\to\infty$.
If $A=0$, the construction provides zero-cost immediate
service for every positive-rate type.
\end{proof}

\section{Proof of the multi-root guarantee}\label{ag:sec:multi-model}

Fixed-root aggregation prescribes one root path for each client.
Here a service instead chooses a root from
a given collection. A client near several roots may be served from
one root in one batch and another in the next. This spatial freedom
belongs to the offline optimum as well as to the original online
problem, so choosing a fixed root for each client requires a proof.

\paragraph{Idea and illustration.}
Imagine two depots at the ends of a tree path. Assigning clients
independently to depots may send neighboring clients in opposite
directions, making a single offline service expensive to reproduce.
We use a tree specialization of the connected $0$-extension construction
of Englert et al.~\cite{connected} to assign connected regions to roots.
Once a region is fixed,
every client in it has one prescribed path to its root. A service of
the free-root optimum can be split at its original time according to
these regions. Every request keeps the same wait; only the spatial
cost can increase. A connected terminal assignment bounds that
increase in expectation, uniformly for every fixed input.

There are three steps. We express activation costs as pendant edges
of a tree. We then apply this connected assignment construction and
compare the restricted and unrestricted offline optima. Finally, the
rooted-forest lemma schedules the resulting fixed paths. The finite construction chooses an assignment once, before arrivals.
Section~\ref{sec:polynomial-roots} replaces its enumeration by a
polynomial local-marginal LP and deterministic support selection.

\subsection{The model and activation costs}
\paragraph{Notation.}

Let $\mathcal T$ be a finite tree with nonnegative edge costs, $R$ a
nonempty set of static facility/root locations, and $c_r\ge0$ their
activation costs. Each positive-rate client vertex $x$ has independent
Poisson arrivals of known rate $\lambda_x$. A service chooses a root $r$,
a connected subtree $S$ containing $r$ and its served requests, and a time
$t$. Its cost is
$$
 c_r+w(S)+\sum_{i\text{ served}}(t-a_i).
$$
For every client location $x$, define its singleton spatial cost
$$
 s_x=\min_{r\in R}\{c_r+d_{\mathcal T}(r,x)\}.
$$
Use the completion and expectation-ratio convention of
Section~\ref{sec:model}. The comparator $\OPT_{\rm free}$ may choose
a different root for every service. A sampled assignment $f$ is
included in the online expectation only.

For each $r$, attach a new terminal $r^+$ to $r$ by a pendant edge of length
$c_r$. A rooted service is now a connected network containing its selected
terminal $r^+$. A nonempty service for ordinary client vertices includes
this pendant edge and pays the activation exactly once. More
precisely, for every nonempty client set $A$,
$$
 w\bigl(\operatorname{hull}_{\widehat T}(r^+,A)\bigr)
 =c_r+w\bigl(\operatorname{hull}_{\mathcal T}(r,A)\bigr).
$$
Thus activation is paid exactly once in each nonempty rooted service.
Contract zero-length edges, merge co-located terminals, and superpose
co-located client streams. These operations preserve every schedule's
cost. Serve zero-singleton-cost types immediately for free and remove
them. Deleting those occurrences from any schedule and adding them
back as free immediate services proves that $\OPT$ is unchanged.
If no type remains, both costs are zero. Otherwise let $K$ be the
remaining terminals. A contracted terminal can be translated back
to any original root in its zero-distance, zero-activation class.

\subsection{Scheduling a rooted forest}\label{sec:forest-scheduling}
Once every client has a prescribed root, the spatial choice disappears,
but the service clocks still have to share common edges. We establish
the scheduling statement needed below, including its offline certificate
and polynomial construction. All arguments in this subsection are local;
no scheduling theorem or numerical certificate from the companion
manuscript is used.

\paragraph{Notation and the certificate.}
Let $F$ be a finite rooted forest with positive rational edge costs
$w_e$. A positive-rate type $i$ occurs at a vertex in one component,
has rational Poisson rate $\lambda_i>0$, and requires the path $P_i$
to that component's root. Zero-length edges are contracted, zero-rate
types are omitted, and types with empty paths are served immediately
for free. Unused edges may be removed. A service buys the union of the
paths of its served requests and pays their individual waiting times.
Unions in different components can be purchased simultaneously as
separate rooted services, without an extra event fee. Thus this
description preserves the original rooted-forest cost exactly.

If no type remains, the algorithm does nothing and both costs are zero.
Otherwise let $\mathcal I$ be the set of remaining types and put
\[
 C_i=w(P_i),\qquad \Gamma=\sum_i\lambda_i,\qquad
 \lambda_{\min}=\min_i\lambda_i,\qquad M=\sum_i\lambda_i C_i.
\]
Fix the rational constants
\[
 r=\frac{101}{100},\qquad \beta=\frac9{10},\qquad s=\frac{11}{5},
 \qquad K_F=\frac{159}{100},\qquad
 \kappa=\frac{593}{900},\qquad C_F=\frac{K_F}{\kappa}=\frac{1431}{593}.
\]
Let $\mathcal A$ consist of consecutive powers of $r$, from the last
power at most $\beta/(2\Gamma)$ to the first power at least
$\max\{2M/\lambda_{\min},\beta/\lambda_{\min}\}$.
Use nonnegative variables $c_{ea}$ and $\delta_{iea}$, where the
latter is present only when $e\in P_i$, and define the rational LP
\begin{align}
 L^*&=\max\sum_{i,e\in P_i,a\in\mathcal A}\lambda_i\delta_{iea},
       \label{eq:forest-lp}\\
 \sum_{e\in P_i,a}\frac{\delta_{iea}}a&\le1,
 &\delta_{iea}&\le c_{ea},\nonumber\\
 a\sum_{i:e\in P_i}\lambda_i\delta_{iea}&\le\beta c_{ea},
 &\sum_a c_{ea}&\le w_e.\nonumber
\end{align}
The first constraint divides a request's waiting budget among the
edges and scales charged to it. The last constraint divides each edge
purchase among scales. The two middle constraints control the size and
mean load of the scalar problems used to certify the offline cost.

\subsubsection{An elementary offline certificate}
We begin with one purchase of cost one and unit waiting cost. Let
$J_H$ be a rate-$u$ Poisson input on $[0,H]$, and let $O_u(H)$ be its
unrestricted offline cost, including completion.

\begin{lemma}[A gap certificate]\label{forest:lem:gap}
For every $u,H\ge0$,
\[
 \E[O_u(H)]\ge(1-e^{-u})H.
\]
\end{lemma}
\begin{proof}
Extend the Poisson process to $H+1$ for the analysis. Charge each
arrival in $[0,H]$ the smaller of one and its gap to the next arrival
of this extended process, using charge one if there is no later arrival.
Consider any batch of any completing
schedule. Its latest arrival has charge at most one, paid by the
batch's purchase. For every other arrival in that batch, the next
arrival of the full process occurs no later than the batch's latest
arrival, hence no later than its service. Its gap charge is therefore
at most its own waiting. Summing over batches proves a pathwise lower
bound, for every realization of the extension. This argument does not
require batches to consist of consecutive arrivals.

At a distinguished arrival at time $a\le H$, the probability of no
arrival in $(a,a+t]$ is $e^{-ut}$ for $0\le t\le1$. These intervals
lie inside $[0,H+1]$, so the expected charge is
$\int_0^1e^{-ut}\,dt$. The Poisson
insertion identity, or conditioning on a distinguished arrival and
integrating its time, therefore gives total expected charge
$uH\int_0^1e^{-ut}\,dt=(1-e^{-u})H$.
The zero-rate case is immediate.
\end{proof}

\begin{lemma}[Forest lower bound]\label{forest:lem:lower}
For every feasible value $L$ in~\eqref{eq:forest-lp} and every $T\ge0$,
\[
 \E[\OPT_T]\ge\kappa LT.
\]
\end{lemma}
\begin{proof}
For each edge--scale pair $(e,a)$, form a weighted scalar problem:
its purchase costs $c_{ea}$, and each original type-$i$ occurrence
requiring $e$ has waiting coefficient $\delta_{iea}/a$.
An original schedule induces a schedule for each scalar problem by
serving occurrences at their original service times. The first and
last LP constraints ensure that the sum of the induced scalar costs
does not exceed the original cost, on every input. Consequently the
original optimum is at least the sum of the scalar optima.

Fix a pair with $c=c_{ea}>0$. Independently retain a type-$i$
occurrence with probability $\delta_{iea}/c$ and give each retained
occurrence waiting coefficient $c/a$. For any fixed weighted scalar
schedule, this preserves expected waiting, while deleting purchases
with no retained request can only decrease purchase cost. Taking the
infimum over the original schedules after averaging the analytical
marks shows that the weighted optimum is at least the expected
retained optimum. The original schedule may be clairvoyant: it is
fixed before these marks are drawn.

Poisson thinning and superposition give retained rate
$\sum_i\lambda_i\delta_{iea}/c$. Normalizing time by $a$ and cost
by $c$ produces a unit-purchase, unit-waiting problem with rate
\[
 u=\frac a c\sum_i\lambda_i\delta_{iea}\le\beta
\]
and horizon $T/a$. Since
$(1-e^{-u})/u=\int_0^1e^{-ut}\,dt$ is nonincreasing in $u>0$,
Lemma~\ref{forest:lem:gap} yields
\[
 \E[O_{ea}]\ge\frac{1-e^{-\beta}}{\beta}
             T\sum_i\lambda_i\delta_{iea}\ge
             \kappa T\sum_i\lambda_i\delta_{iea}.
\]
For the last inequality it suffices to use the elementary rational
comparison
\[
 e^{9/10}>\sum_{j=0}^{5}\frac{(9/10)^j}{j!}
       =\frac{9835033}{4000000}>\frac{1000}{407}.
\]
Thus $(1-e^{-\beta})/\beta>593/900$. Pairs with $c=0$ or $u=0$
contribute zero. Summing proves the assertion with no horizon loss.
\end{proof}

\subsubsection{From the dual to synchronized clocks}
The dual of~\eqref{eq:forest-lp} minimizes
$\sum_i y_i+\sum_e w_ez_e$ over nonnegative variables satisfying
\[
 \frac{y_i}{a}+b_{iea}+a\lambda_i t_{ea}\ge\lambda_i,
 \qquad z_e\ge\beta t_{ea}+\sum_{i:e\in P_i}b_{iea}.
\]
Fix an optimal dual solution and put $\tau_i=y_i/\lambda_i$.
The feasible solution $y_i=t_{ea}=0$, $b_{iea}=\lambda_i$,
$z_e=\Lambda_e:=\sum_{i:e\in P_i}\lambda_i$ has value $M$.
Hence strong duality gives
\[
 \sum_i y_i+\sum_e w_ez_e=L^*\le M,
 \qquad 0\le\tau_i\le M/\lambda_{\min}.
\]
For each edge, sort its users by $\tau_i$, giving type $i$ an interval
of length $\lambda_i$. Denote the resulting nondecreasing step function
by $\tau_e(v)$, $0<v<\Lambda_e$, and put
\[
 G_e(v)=\int_0^v\tau_e(u)\,du,\qquad
 Z_e=\max_{0\le v\le\Lambda_e}v\bigl(1-G_e(v)/\beta\bigr).
\]

\begin{lemma}[Prefix and supporting-line bounds]\label{forest:lem:prefix}
For every used edge, $0<Z_e\le r z_e$. There is $h_e\in[1,2]$
such that, for every $q>0$,
\[
 \int_0^{\Lambda_e}e^{-qG_e(v)}\,dv
 \le Z_e\left(2h_e-h_e^2+\frac{h_e^2}{q\beta}\right).
\]
\end{lemma}
\begin{proof}
For a grid scale $a$ with $v=\beta/a\le\Lambda_e$, take the first
$v$ units of rate in the sorted list, using fractions $\alpha_i\in[0,1]$
if the prefix ends inside a type. Thus
$\sum_i\alpha_i\lambda_i=v$ and
$\sum_i\alpha_i\lambda_i\tau_i=G_e(v)$. Summing the dual
constraints with these fractions and using $av=\beta$ gives
\[
 z_e\ge\beta t_{ea}+\sum_i b_{iea}
 \ge v-G_e(v)/a=v\bigl(1-G_e(v)/\beta\bigr).
\]
Let $v_*>0$ maximize this expression; its maximum is positive because
$G_e(v)\to0$ as $v\downarrow0$. If
$\tau_{\max,e}=\max_{i:e\in P_i}\tau_i>0$, the derivative of
$v(1-G_e(v)/\beta)$, where it exists, is at least
$1-2v\tau_{\max,e}/\beta$. A maximizer can therefore be chosen with
$v_*\ge\min\{\Lambda_e,\beta/(2\tau_{\max,e})\}$.
If $\tau_{\max,e}=0$, choose $v_*=\Lambda_e$.
It follows that $a_*=\beta/v_*$ lies between the LP grid endpoints.
Round it upward to a grid point $a'<r a_*$ and set $v'=\beta/a'$.
The grid inequality and monotonicity of $G_e$ imply
$z_e\ge v'(1-G_e(v')/\beta)\ge Z_e/r$.

Write $G=G_e$, $Z=Z_e$, and set
$g=(\beta-G(v_*))/v_*=\beta Z/v_*^2>0$.
The left and right derivative conditions at a maximizer show that
$g$ is a supporting slope for the convex function $G$. At the right
endpoint $v_*=\Lambda_e$, the left derivative condition gives
$g\ge G'_-(v_*)$, which is again a supporting slope on this domain.
Thus $G(v)\ge G(v_*)+g(v-v_*)$. Applying this inequality at zero
gives $G(v_*)\le\beta/2$. Consequently
$h_e=v_*/Z=1/(1-G(v_*)/\beta)$ lies in $[1,2]$, and
\[
 G(v)\ge\beta\max\left\{0,
       \frac{v-Z(2h_e-h_e^2)}{Zh_e^2}\right\}.
\]
Integrating the exponential of this lower bound, extending its
nonnegative affine part to $[0,\infty)$, proves the assertion.
\end{proof}

For a common phase $\theta\in[0,1)$, serve zero-$\tau_i$ types
immediately and assign every other type period
\[
 p_i(\theta)=s2^{\theta+\lceil\log_2\tau_i-\theta\rceil}.
\]
All clocks have time origin zero. At a tick, clear exactly the due nonempty
type queues, buying their path union in each rooted component.
Positive periods are integer powers of two times a common factor,
so slower ticks are also faster ticks. Every request is served by its
next own tick, and the policy completes finite inputs without an end signal.

For an arbitrary vector of such nested periods $p$, sort users of
edge $e$ by period, write $p_e(v)$ for their rate-quantile function,
and define
\[
 P_e(v)=\int_0^v p_e(u)\,du,\qquad
 R_e(p)=\int_0^{\Lambda_e}e^{-P_e(v)}\,dv,\qquad
 Q(p)=\frac12\sum_i\lambda_i p_i+\sum_e w_eR_e(p).
\]
Zero periods designate immediate service.

\begin{lemma}[Complete clock cost]\label{forest:lem:clock-cost}
For every fixed nested-period vector and every $T\ge0$,
\[
 \E[\ALG_T]\le TQ(p)+B_F(p),\qquad
 B_F(p)=\sum_i C_i+\frac18\sum_i\lambda_i p_i^2.
\]
\end{lemma}
\begin{proof}
Fix an edge. Let $\lambda_0$ be its immediate rate, and let
$p_1<\cdots<p_m$ be its distinct positive periods with group rates
$\gamma_j$. Put $A_0=0$ and $A_j=\sum_{k\le j}\gamma_kp_k$.
A tick of $p_j$ that is not a tick of $p_{j+1}$ has precisely groups
$1,\ldots,j$ due. Their pending arrivals occupy independent intervals
of lengths $p_k$, so its nonempty probability is $1-e^{-A_j}$.
Nesting and summation by parts give expected purchases through $T$
equal to
\[
 \lambda_0T+\sum_{j=1}^m\left\lfloor\frac T{p_j}\right\rfloor
                      (e^{-A_{j-1}}-e^{-A_j})
 \le T\left[\lambda_0+
       \sum_{j=1}^m\frac{e^{-A_{j-1}}-e^{-A_j}}{p_j}\right]
 =TR_e(p).
\]
Immediate arrivals meet deterministic ticks only on a null event.
The formula also covers $m=0$. After $T$, each positive-period type
has at most one final clearing, with total spatial cost at most
$\sum_i C_i$. A type-$i$ arrival waits at most until its next own
tick. Integrating that residual waiting over $[0,T]$ gives
$\lambda_i(Tp_i/2+p_i^2/8)$: the excess on a final partial period
is at most $\lambda_i p_i^2/8$. Summing proves the claim.
\end{proof}

\begin{lemma}[An averaged clock bound]\label{forest:lem:phase-average}
For a uniform phase $\theta\in[0,1)$ and $\bar K=15891/10000$,
\[
 \E_\theta[Q(p(\theta))]\le\bar K L^*<K_F L^*.
\]
\end{lemma}
\begin{proof}
Write $\ell=\ln2$. For each positive $\tau$, the ratio
$p_\theta(\tau)/(s\tau)$ has distribution $2^U$ for uniform
$U\in[0,1)$. In particular the mean waiting rate is
$[s/(2\ell)]\sum_i y_i$. The multipliers for different types need
not be independent.

Fix $v$ with $G_e(v)>0$. Jensen's inequality with weights
$\tau_e(u)/G_e(v)$ gives
\[
 \E_\theta[e^{-P_e(v)}]
 \le\E_U[e^{-s2^U G_e(v)}].
\]
Indeed, before averaging, apply convexity of the exponential to the
weighted mean of the multipliers, and then use their identical
one-dimensional distributions. Zero-$\tau$ terms have zero weight.
For $G_e(v)=0$ both sides equal one. Tonelli's theorem and
Lemma~\ref{forest:lem:prefix} now yield, with the same $h=h_e$
for all multipliers,
\[
 \E_\theta[R_e(p(\theta))]
 \le Z_e\left[2h-\left(1-\frac1{2s\beta\ell}\right)h^2\right]
 \le\frac{r z_e}{1-1/(2s\beta\ell)}.
\]
The last step completes the square. Its denominator is positive.
Both coefficients needed to bound the dual objective satisfy
\[
 \frac{s}{2\ell}<\frac{1100}{693}<\bar K,\qquad
 \frac{r}{1-1/(2s\beta\ell)}<\frac{6929307}{4360700}<\bar K.
\]
These are elementary comparisons: the first three positive terms of
the logarithm series give
\[
 \ell>2\left(\frac13+\frac1{81}+\frac1{1215}\right)
       =\frac{842}{1215}
       >\frac{693}{1000},
\]
which implies the displayed inequalities for our fixed rational
$r,\beta,s$. The series follows by integrating
$2/(1-t^2)$ from zero to $1/3$. Combining the waiting and purchase
bounds with the dual objective proves the lemma. Positivity of $L^*$
is verified below. The gap between $\bar K$ and $K_F$ will pay for
the deterministic phase selection.
\end{proof}

\subsubsection{Polynomial deterministic choice of rational periods}
All preprocessing below uses the known forest and rates. Let $B$ be
their total binary encoding length. The scale grid has polynomial size and
encoding length: its endpoints have logarithmic magnitudes polynomial
in $B$, and $r,\beta$ are fixed rationals. Exact comparisons of rational
powers locate the endpoints. The primal and dual LPs therefore have
polynomially many rational variables and constraints, and rational
linear programming supplies optimal solutions of polynomial encoding
length in polynomial time.

Moreover $L^*>0$ in every nontrivial instance. Choose one incidence
$(i,e)$ and one scale $a$, set $c_{ea}=w_e$ and
$\delta_{iea}=\min\{w_e,a,\beta w_e/(a\lambda_i)\}>0$, and set
all other variables to zero. This is a feasible positive solution.
In particular, accuracy proportional to $L^*$ requires only
polynomially many bits.

Replace $\theta$ by $b=2^\theta\in[1,2)$ and write
\[
 k_i(b)=\min\{k\in\mathbb Z:b2^k\ge\tau_i\},\qquad
 p_i(b)=sb2^{k_i(b)}
\]
for positive-$\tau_i$ types, with zero otherwise. Each such type has
one possible breakpoint, the rational mantissa
$b_i=\tau_i/2^{\lfloor\log_2\tau_i\rfloor}\in[1,2)$.
Binary exponents and mantissas are computed by exact rational
comparisons. At a breakpoint the exponent drops, and its value
belongs to the cell on its right.

Set $\eta=1/3000$ and use the finite rational candidate set
\[
 \mathcal H=\{1+j\eta:j\in\mathbb Z_{\ge0},\ 1+j\eta<2\}
       \cup\{b_i:\tau_i>0\}.
\]
It has at most $|\mathcal I|+3000$ members of polynomial encoding length.
The following scaling observation controls its approximation error.
For $0<c\le1$, convexity of $P_e$ and $P_e(0)=0$ give
\[
 cR_e(cp)=\int_0^{c\Lambda_e}e^{-cP_e(u/c)}\,du
       \le\int_0^{c\Lambda_e}e^{-P_e(u)}\,du\le R_e(p).
\]
Hence $Q(cp)\le c^{-1}Q(p)$, including immediate types.
Choose $b_*\in[1,2)$ supplied by
Lemma~\ref{forest:lem:phase-average}, and let $b\in\mathcal H$
be its largest predecessor. Either $b=b_*$, or there is no breakpoint
in $(b,b_*]$. Thus $p(b)=c p(b_*)$ with
$c=b/b_*\ge1/(1+\eta)$, and some candidate satisfies
\[
 Q(p(b))\le(1+\eta)\bar K L^*.
\]
All candidate periods are rational and nested; no logarithm has to
be evaluated to make a rounding decision.

For each candidate, compute an enclosing interval for $Q$ of width
at most $\eta L^*$ and choose the candidate with smallest upper
endpoint. Its true value is at most
\[
 (1+\eta)\bar K L^*+\eta L^*
 =\frac{47698891}{30000000}L^*<K_F L^*.
\]
To see that certified evaluation is polynomial, use the finite sum
for $R_e$ in Lemma~\ref{forest:lem:clock-cost}. All its exponential
arguments and coefficients are rational of polynomial encoding
length. The sum of absolute coefficient magnitudes and the positive
rational tolerance $\eta L^*$ determine a polynomial number of
accuracy bits; no small difference requires relative accuracy.
For a nonnegative rational $x$, $e^{-x}$ can be enclosed to absolute
error $2^{-b}$ in polynomial time. If $x\ge b+1$, the interval
$[0,2^{-b-1}]$ suffices. Otherwise $x$ has polynomial magnitude,
and the Taylor expansion at zero, with remainder bounded by
$x^{N+1}/(N+1)!$, reaches the required accuracy with polynomially
many rational operations. Interval arithmetic gives the stated
candidate enclosures. This is an algorithm for certified real
evaluation, not a separate computational assumption or a finite
certificate for a theorem constant.

\begin{theorem}[Independent rooted-forest scheduling]\label{thm:forest-scheduling}
For every nontrivial rational rooted-forest instance, the certificate
$L^*$ in~\eqref{eq:forest-lp} is positive and satisfies, for all $T\ge0$,
\[
 \E[\OPT_T]\ge\kappa L^*T.
\]
Deterministic preprocessing polynomial in the binary input length produces
rational nested periods and a causal completing policy such that
\[
 \E[\ALG_T]\le K_F L^*T+B_F(p).
\]
Consequently its long-run ratio of expectations is at most
$C_F=1431/593$. The all-free instance
is served immediately at zero cost.
\end{theorem}
\begin{proof}
The lower estimate is Lemma~\ref{forest:lem:lower}. The deterministic
phase selection satisfies the stated upper rate bound, and
Lemma~\ref{forest:lem:clock-cost} includes every final clearing and
waiting cost. Each positive period is finite, so completion needs
no end signal. Positivity of $L^*$ makes the offline expectation
grow linearly on a fixed nontrivial instance, while $B_F(p)$ is
independent of $T$. Division and passage to the limit give the ratio.
\end{proof}

\subsection{Choosing connected root regions}
A terminal retraction $f:V(\widehat T)\to K$ fixes every terminal.
We require each fiber to be connected, so that the path from a client
to its assigned terminal stays inside one region. We obtain these
regions by specializing the growing construction of Englert
et al.~\cite[Section~4, Algorithm~1]{connected} to the tree.
The following comparison is uniform over the offline service and
its client set.

\begin{theorem}[Connected terminal assignment]\label{ag:mr:thm:structure}
There is a finitely supported distribution of connected terminal
retractions such that every connected service subtree $S$ containing
a terminal and every client set $A\subseteq V(S)$ satisfy
$$
 \E_f\left[\sum_{r\in K}w\!\left(\bigcup_{x\in A:f(x)=r}P_{\widehat T}(x,r)\right)\right]
 \le\frac{313}{8}w(S).
$$
Consequently, for every finite input,
$\E_f[\OPT_f(I)]\le(313/8)\OPT_{\rm free}(I)$.
\end{theorem}
\begin{proof}
Appendix~\ref{app:partition} gives this tree specialization with a finite
mixture of quadrupling scales and proves the service inequality.
Split each free-root offline batch by assigned terminal at its
original service time. The displayed inequality controls its expected
spatial cost, while every occurrence keeps the same waiting time.
Taking the infimum over original schedules proves the second claim.
\end{proof}
The construction freezes assigned vertices and attaches each new
connected component to one already assigned neighbor. Its labels
never change. A random initial scale lets the analysis charge the
first and last endpoint assignments jointly. The distribution is
chosen once from the spatial instance, before any requests arrive.
\subsection{From connected regions to an online algorithm}\label{ag:sec:multi-main}

Fix a connected retraction $f$ produced by
Theorem~\ref{ag:mr:thm:structure}. For each positive-rate client type
$x$, define its required root path
\begin{equation}\label{ag:eq:mr-required}
 R_x^f=P_{\widehat T}(x,f(x)),
 \qquad C_x^f=w(R_x^f).
\end{equation}
Every such path lies in the connected fiber $f^{-1}(f(x))$, because a
connected subset of a tree contains the unique path between any two
of its vertices. If $f(x)\ne f(y)$, the two paths in
\eqref{ag:eq:mr-required} are edge-disjoint. The pendant edge of a
terminal belongs to every nonempty path assigned to that terminal,
so purchasing a union of paths pays its activation cost exactly once.

Delete every edge whose endpoints have different labels under $f$.
The resulting forest has one component for each nonempty fiber,
rooted at its terminal: connectedness makes the fiber a tree, and a
second terminal cannot lie in it because terminals keep their own labels.
Remove components and branches with no positive-rate clients. Denote
the remaining rooted forest by $F_f$.

\begin{lemma}[Exact rooted-forest reduction]
\label{ag:lem:mr-resource-equality}
For every finite input $I$ and every connected terminal retraction $f$,
the rooted-forest optimum on $F_f$ equals $\OPT_f(I)$, the multi-root
optimum restricted to $f$.
\end{lemma}
\begin{proof}
Trim each $f$-restricted service to the union of its clients' paths to
their assigned terminal. This is a feasible purchase in $F_f$ with
no greater cost. Conversely, split any forest purchase by component
at its original service time. Each nonempty path union is connected
to its assigned terminal and translates back into a valid rooted
service. Components have disjoint edges, and the pendant activation
edge is paid exactly once for each such service. Zero-cost contracted
activation edges add nothing. Thus splitting preserves both spatial
cost and waiting, proving equality.
\end{proof}

For each assignment, apply Theorem~\ref{thm:forest-scheduling}.
Write $L_f^*$ for the rooted-forest LP value
and $B_f$ for the completion constant of its finite-preprocessing
policy. The preceding exact reduction gives
\[
 \E_I[\ALG_f(I_T)]\le K_F L_f^*T+B_f,
 \qquad \E_I[\OPT_f(I_T)]\ge\kappa L_f^*T.
\]
At each tick, split the due requests by assigned terminal and buy
the union of their assigned paths as one rooted service per terminal.
Connected fibers make these path unions edge-disjoint, so the
forest purchase cost is reproduced exactly.

\begin{theorem}[Finite-preprocessing multi-root guarantee]
\label{ag:thm:mr-final}
For rational costs and rates, the algorithm with one assignment sampled
before arrivals satisfies
\[
 \E_{f,I}[\ALG_f(I_T)]
 \le\frac{313}{8}C_F\,\E_I[\OPT_{\rm free}(I_T)]+\E_f[B_f].
\]
Its long-run ratio is at most $(313/8)C_F=447903/4744<94.415$.
A deterministic algorithm with the same bound is obtained by
finite exhaustive preprocessing over connected assignments.
\end{theorem}
\begin{proof}
Combine the fixed-assignment inequalities with
Theorem~\ref{ag:mr:thm:structure}. There are finitely many assignments,
each with finite periods, so $\E_f[B_f]$ is finite. On the nonfree
instance $L_f^*>0$ for every assignment. The same inequalities give
$\E[\OPT_{\rm free}(I_T)]\ge(8\kappa/313)T\,\E_f[L_f^*]>0$,
which justifies the long-run division.

For deterministic selection choose $f^*$ minimizing $L_f^*$ among
all connected assignments. This is finite preprocessing and uses
only the known instance. For every $T>0$,
\[
 L_{f^*}^*\le\E_f[L_f^*]
 \le\frac{1}{\kappa T}\E_{f,I}[\OPT_f(I_T)]
 \le\frac{313}{8\kappa T}\E_I[\OPT_{\rm free}(I_T)].
\]
Substitution in the online estimate proves the same finite-horizon
bound with $B_{f^*}$.
\end{proof}

The exhaustive selection in this theorem is finite, without a
polynomial running-time assertion. The next section replaces it
by a polynomial construction and uses the rational-period forest
policy with the same competitive factor.

\subsection{Polynomial deterministic root preprocessing}\label{sec:polynomial-roots}

Assume binary-encoded rational tree lengths, root activation costs,
and positive arrival rates. Zero-length contractions and free types
are handled as in Section~\ref{ag:sec:multi-model}. Let $B$ denote the total binary input length. The all-free instance
is served immediately and bypasses the LPs below. The nontrivial
case uses only rational linear programming and the local
rooted-forest scheduling lemma; the original partition distribution is needed
to prove feasibility, not to generate samples.

Let $T=(V,E)$ be the positive-length augmented tree after the
activation-edge construction of Section~\ref{ag:sec:multi-model} and zero-length contractions. Let
$\mathcal K\subseteq V$ be its nonempty set of terminals,
$n=|V|$, and $k=|\mathcal K|$. Fix an arbitrary orientation from a
vertex $o$. For a child $v$, let $T_v$ denote its descendant subtree
and $\mathcal K_v=\mathcal K\cap V(T_v)$.

For a directed edge $uv$ with parent $u$ and child $v$, allow the label
pair $(a,b)\in\mathcal K^2$ exactly when
\begin{equation}\label{eq:allowed}
 a=b\quad\hbox{or}\quad
 [a\notin\mathcal K_v\ \hbox{and}\ b\in\mathcal K_v].
\end{equation}
Terminals must have their own labels.

\begin{lemma}[Local characterization]\label{lem:local}
A labeling $f:V\to\mathcal K$ fixing terminals is a connected terminal
retraction if and only if every edge satisfies \eqref{eq:allowed}.
\end{lemma}
\begin{proof}
For a connected retraction, an edge between distinct fibers separates
their terminals: otherwise the path from an endpoint to its own
terminal would cross the edge and enter the other fiber. Thus unequal
labels lie on the specified opposite sides.

Conversely, let $f(x)=a$ and follow the path from $x$ to terminal $a$.
At any edge on this path, the side containing the current vertex
does not contain terminal $a$. A transition to a different label
would require, by \eqref{eq:allowed}, that the terminal of the current
label lie on its current side. This is impossible. Inductively every
vertex on the path has label $a$. Hence each fiber contains its
terminal and is connected.
\end{proof}

Define the rational edge cost for an allowed pair by
\[
 W_{uv}(a,b)=
 \begin{cases}
  \ell_{uv},&a=b,\\
  d_T(a,b),&a\ne b.
 \end{cases}
\]
This is precisely the mixed edge cost $W_f(e)$ in the deterministic
service conversion of Appendix~\ref{app:partition}.

Introduce vertex marginals $x_{v,a}\ge0$ and allowed edge marginals
$y_{uv,a,b}\ge0$. Disallowed edge entries are zero. Impose
\begin{align}
 \sum_a x_{v,a}&=1 &&(v\in V),\label{eq:vertex}\\
 x_{a,a}&=1 &&(a\in\mathcal K),\label{eq:terminal}\\
 \sum_b y_{uv,a,b}&=x_{u,a},&
 \sum_a y_{uv,a,b}&=x_{v,b} &&(uv\in E),\label{eq:consistency}\\
 \sum_{a,b}W_{uv}(a,b)y_{uv,a,b}
        &\le \frac{313}{8}\ell_{uv} &&(uv\in E).\label{eq:stretch}
\end{align}
There are $O(nk^2)$ variables and polynomially many constraints,
with rational coefficients of polynomial encoding length.

\begin{lemma}\label{lem:feasible}
The LP \eqref{eq:vertex}--\eqref{eq:stretch} is feasible and an exact
rational feasible point can be obtained in polynomial time.
\end{lemma}
\begin{proof}
The construction in Appendix~\ref{app:partition} proves the
\emph{edgewise} statement
$\E_f[W_f(e)]\le(313/8)\ell_e$ for every edge, including terminal edges.
Its vertex and edge marginals satisfy every displayed constraint by
Lemma~\ref{lem:local}. This proves feasibility, irrespective of whether
that original distribution has algebraic probabilities.
The present LP has rational data and polynomial size, so rational
linear programming supplies a polynomial-size rational feasible point.
No sampling from the original construction is needed.
\end{proof}

\subsubsection{An explicit polynomial-support decomposition}

\begin{lemma}\label{lem:decompose}
Any rational point satisfying \eqref{eq:vertex}--\eqref{eq:consistency}
can be decomposed in polynomial time into at most
$nk+(n-1)k^2$ connected terminal retractions with nonnegative rational
weights summing to one and exactly the given vertex and edge marginals.
\end{lemma}
\begin{proof}
Keep residual marginals initially equal to $x,y$. At each step they
have a common total mass $m>0$ at every vertex, obey consistency,
and retain all forbidden zeros. Choose any positive label at $o$.
Proceed down the tree. If parent $u$ has chosen label $a$, its positive
residual $x_{u,a}$ is the sum of row $(a,\cdot)$ at each child edge.
Choose any positive entry $y_{uv,a,b}$ in that row and give $v$ label $b$.
Column consistency makes the selected $x_{v,b}$ positive as well.

The resulting labeling fixes terminals and uses only allowed pairs,
so it is connected by Lemma~\ref{lem:local}. Let $\zeta>0$ be the
smallest selected vertex or edge entry. Record this labeling with
weight $\zeta$, and subtract $\zeta$ from all its selected vertex
and edge entries.

Nonnegativity is preserved. Each selected vertex entry and each
corresponding selected row or column decrease by the same $\zeta$,
so consistency remains true, with common mass $m-\zeta$.
At least one positive entry vanishes; none can reappear. Consequently
there are at most as many iterations as initial vertex and edge
coordinates. At zero common mass every residual entry is zero,
proving exact reconstruction and that the recorded weights sum to one.

The arithmetic is polynomial. All residual values are differences of
initial rational entries and previous subtraction weights; a common
denominator given by the product of their initial denominators has
polynomial bit length. The number of iterations is polynomial.
\end{proof}

The residual need not satisfy \eqref{eq:stretch}; only the original
point needs those constraints. Its exactly reconstructed mixture
therefore satisfies the required expected stretch. Individual support
assignments need not satisfy the edgewise bound separately.

\subsubsection{Deterministic selection from the support}

Let $(\zeta_j,f_j)_{j=1}^N$ be the preceding decomposition.
For every fixed connected free-root service $S$ and client set
$A\subseteq V(S)$, the deterministic conversion in Appendix~\ref{app:partition} gives
\[
 \sum_{a\in\mathcal K}
 w\!\left(\bigcup_{x\in A:f_j(x)=a}P_T(x,a)\right)
 \le\sum_{e\in E(S)} W_{f_j}(e).
\]
Averaging and using \eqref{eq:stretch} yields the same
$313/8$ service bound as before. Splitting a free-root offline schedule
by labels at its original times preserves all waiting. Since
$313/8\ge1$, it follows for every finite input $I$ that
\begin{equation}\label{eq:mix-opt}
 \sum_j\zeta_j\OPT_{f_j}(I)\le
                  \frac{313}{8}\OPT_{\mathrm{free}}(I).
\end{equation}

For each $f_j$, form its rooted forest as above and solve the local LP;
write its value as $L_j^*$. There are only polynomially many such LPs.
Every path incidence description and every LP grid has polynomial
encoding size. Choose $j_*$ minimizing the rational $L_j^*$, breaking
ties by a fixed rule. This choice uses only the known instance.

For every $T>0$, the forest certificate and \eqref{eq:mix-opt} imply
\[
 \kappa L_{j_*}^*T
 \le\kappa T\sum_j\zeta_jL_j^*
 \le\sum_j\zeta_j\,\E[\OPT_{f_j,T}]
 \le\frac{313}{8}\,\E[\OPT_{\mathrm{free},T}].
\]
Run the polynomial forest preprocessing only for this selected
assignment. Its rate bound gives
\[
 \E[\ALG_T]
 \le K_F L_{j_*}^*T+K''_I
 \le\frac{313}{8}C_F\,
                       \E[\OPT_{\mathrm{free},T}]+K''_I.
\]
The $T=0$ case is immediate. Connected fibers make the corresponding
path unions edge-disjoint, so the forest policy translates back into
valid rooted services without additional cost. It completes every
finite input without an end signal. This proves the following guarantee.

\begin{theorem}[Polynomial multi-root guarantee]
\label{thm:efficient-roots}
For every rational multi-root tree instance, deterministic preprocessing
polynomial in the binary input length $B$ produces one fixed connected assignment
and a causal policy completing every finite input without an end
signal such that
\[
 \E[\ALG_T]\le\frac{313}{8}C_F\,
                   \E[\OPT_{\rm free,T}]+B_I
 \qquad(T\ge0),
\]
where $B_I$ is finite and independent of $T$.
The long-run ratio is at most $447903/4744<94.415$.
\end{theorem}
\begin{proof}
The preceding construction gives the finite-horizon inequality,
using Theorem~\ref{thm:forest-scheduling} for the selected assignment.
There are $O(nk^2)$ marginal coordinates and at most
$nk+(n-1)k^2$ support assignments. Each prescribed-path instance
and its rooted-forest LP have polynomial rational encoding size.
The marginal LP, its constructive decomposition, and all support
certificate LPs therefore take polynomial time. Only the chosen
assignment needs the polynomial clock preprocessing.

Every nonfree assignment has $L_f^*>0$ by
Theorem~\ref{thm:forest-scheduling}. Thus the averaged certificate
comparison above gives a positive linear lower bound for
$\E[\OPT_{\rm free,T}]$, and the finite completion term vanishes
after division as $T\to\infty$. The numerical bounds follow from
$C_F=1431/593$. The all-free case has zero cost throughout.
\end{proof}

\paragraph{Scope of the polynomial guarantee.}
The $O(nk^2)$ bound is the size of the marginal LP, not a bound on
the full running time. Polynomiality is in the binary input length
and is not strong polynomiality. Connected
fibers and exact marginal decomposition use the unique paths of a
tree; no analogous general-graph statement is asserted. The support
assignments need not each obey the edgewise stretch constraints.
Their mixture does, and the smallest forest certificate is what
permits deterministic selection.

\section{Proof of the tree-service guarantee}\label{sec:mobile}
In online service, the spatial decision is a trajectory. A server
serves pending requests by visiting their locations and retains
its final position for the next service. This distinguishes it
from a network that must be bought again at every event. In
particular, a lower bound based only on distance from a designated
root need not hold for a server that stays near the current demand.

\paragraph{Idea and illustration.}
Consider two sites separated by a long edge. If only the far site
receives requests, the server can move there once and stay. A
policy forced to return home after every service would repeatedly
pay an expense the optimum avoids. If both sites receive requests,
however, avoiding a crossing leaves demand on one side waiting.
The rates on both sides, rather than demand on just the far side,
are therefore the appropriate basis for a lower bound.

On a tree we choose a rate-weighted centroid as the online home:
each branch contains at most half the total rate. This ensures
enough demand outside a branch to compare with its native stream.
Independent marks assign each occurrence at most one auxiliary target.
A second thinning distributes its waiting budget over one two-site
problem per edge. A telescoping sum pays for all these projections,
even though the offline trajectory need not return home.

\subsection{The mobile-service model}\label{os:sec:prelim}

\paragraph{Notation.}
Let $\mathcal T=(V,E)$ be a finite tree with positive edge lengths
$\ell_e$. Zero-length edges can be contracted. A server starts at a
specified vertex $s_0$. A move from $u$ to $v$ costs $d(u,v)$ and takes
zero time; intermediate vertices on the unique path are visited. A visit
serves all pending requests at that vertex at no additional service fee.
Requests at the server's current vertex are served immediately. The server
may move without a pending request, and the offline comparator may do so
proactively.

For every vertex $v$, an independent homogeneous Poisson process of
known rate $\lambda_v\ge0$ generates requests. A request $i$ arrives at
vertex $v_i$ at time $a_i$. If the server first serves it at $t_i\ge a_i$,
its delay cost is $t_i-a_i$. The schedule's cost is total movement plus
$\sum_i(t_i-a_i)$. Requests are distinguishable even when several occur
at the same vertex. The algorithm sees arrivals as they happen and never
sees future arrivals.

Let $I_T$ contain arrivals in $[0,T]$, and let $\ALG_T$ be the complete
cost of our algorithm on this finite input. It runs without an
end-of-input signal and must eventually clear every request; movements
and waiting after $T$ needed to do so are included. The offline optimum
$\OPT_T$ knows $I_T$, may choose any feasible trajectory, and is not
required to return to any fixed point. We prove an inequality of the form
\begin{equation}\label{os:eq:finite-contract}
 \E[\ALG_T]\le C\,\E[\OPT_T]+K(\mathcal T,\lambda,s_0),
\end{equation}
where $K$ is finite and independent of $T$. For an instance with positive
optimal rate, this implies the long-run ratio of expectations
$\limsup_{T\to\infty}\,\E[\ALG_T]/\E[\OPT_T]\le C$. We do not replace it by
the expectation of a samplewise ratio.

At a common time, arrivals are observed before service actions, and
simultaneous moves are processed in execution order. Charging intervals
are half-open: an action at the left endpoint belongs to that interval,
and one at the right endpoint belongs to the next.

\subsection{A two-site tool for the tree lower bound}\label{os:sec:two-point}

Consider two sites at distance $w>0$.
Each request has waiting coefficient $\rho>0$.
The online and offline servers start at the same specified endpoint.
While at one endpoint, serve arrivals there immediately.
Wait until the total weighted waiting accumulated by the pending
requests at the opposite endpoint reaches $w$, then cross and clear
all of them.
An empty queue triggers no action.

\begin{lemma}\label{lem:r5-two-pathwise}
For every finite request input, the threshold-crossing policy completes
the input without an end signal and satisfies
$$
 \operatorname{ALG}(I)\le4\operatorname{OPT}(I).
$$
The assertion permits arbitrary arrival times and does not require a
Poisson input.
\end{lemma}
\begin{proof}
Call the interval from one online crossing to the next a phase, with
the first phase starting at time zero.
At the beginning of every phase the opposite queue is empty.
Every completed phase incurs movement $w$ and weighted waiting $w$,
so $n$ online crossings cost exactly $2nw$.
If a finite input leaves a nonempty opposite queue, its waiting grows
until the threshold is reached.
Thus every pending request is eventually served, and there are only
finitely many completed phases.

The first phase alone forces an offline payment of at least $w$.
If the offline trajectory crosses in this phase, charge a crossing.
Otherwise it stays at the common initial endpoint, so the requests
in the online opposite queue incur at least the entire online
phase's weighted waiting, namely $w$.

Pair the remaining phases as $(2,3),(4,5),\ldots$.
In each complete pair, either the offline trajectory crosses, paying
at least $w$, or it stays at one endpoint throughout the pair.
The two online opposite queues are at different endpoints.
One phase's opposite requests are therefore unvisited throughout
their phase and incur at least its weighted waiting $w$.
All charging intervals are half-open.
A crossing at a right endpoint belongs to the next interval; if it
is the first visit serving the preceding requests, those requests
have already incurred the claimed waiting.
Movement charges use disjoint time intervals, and delay charges use
disjoint sets of requests classified by their online arrival phase.
They therefore do not reuse either cost account.

For $n\ge1$, this gives
$\operatorname{OPT}(I)\ge\lceil n/2\rceil w$.
Consequently $2nw\le4\operatorname{OPT}(I)$.
The case $n=0$ has zero online cost.
All intervals, movements, and waiting used here may extend beyond
the arrival horizon, as required by the completion convention.
\end{proof}

If the two initial endpoints were allowed to differ, pairing phases
from the first would still give
$\operatorname{ALG}\le4\operatorname{OPT}+2w$.
We use the stronger same-initial-endpoint statement: each virtual
cut inherits its initial side from the given original position $s_0$.
This fixes no home or subsequent trajectory for the offline server.

Write
$$
 g(a)=\min_{n\ge1}\left\{a/n+(n-1)/2\right\}.
$$
The following elementary potential is the only scalar batching fact
needed in the service analysis. We include its proof so that the
two-site certificate is independent of any scalar or aggregation model.

\begin{lemma}[Bounded queue potential]\label{lem:service-scalar-potential}
For $a>0$, choose an integer $N\ge1$ with
$N(N-1)/2\le a\le N(N+1)/2$. Then
$g(a)=a/N+(N-1)/2$ and $N-1\le g(a)\le N$. The function
\[
 h_a(n)=
 \begin{cases}
 ng(a)-n(n-1)/2,&0\le n\le N,\\
 a,&n\ge N
 \end{cases}
\]
is nondecreasing, satisfies $h_a(0)=0$ and $0\le h_a(n)\le a$,
and obeys
\[
 n+h_a(n+1)-h_a(n)\ge g(a)\qquad(n\ge0).
\]
\end{lemma}
\begin{proof}
Consecutive terms $a/n+(n-1)/2$ in the definition of $g$ differ by
$1/2-a/[n(n+1)]$. Thus $N$ is a minimizer and the stated bounds on
$g(a)$ follow from the inequalities defining $N$. The two formulas for
$h_a$ agree at $N$. For $n<N$, its increment is $g(a)-n\ge0$, and
$n+h_a(n+1)-h_a(n)=g(a)$. For $n\ge N$, the increment is zero and
$n\ge N\ge g(a)$. This proves all assertions, including at a common
endpoint where either adjacent choice of $N$ is valid.
\end{proof}

This potential supplies the following two-sided Poisson lower certificate.

\begin{lemma}\label{lem:r5-two-lower}
Suppose the two endpoints receive independent Poisson streams of the
same rate $\mu>0$, their distance is $w>0$, and requests have waiting
coefficient $\rho>0$.
For either prescribed initial endpoint and every $T\ge0$,
$$
 \mathbb E[\operatorname{OPT}^{\rm two}_T]
 \ge\frac{\rho}{4}g(\mu w/\rho)T.
$$
\end{lemma}
\begin{proof}
Apply Lemma~\ref{lem:service-scalar-potential} to the threshold policy,
with $a=\mu w/\rho$. If $n$ requests wait opposite the server, put
$V(n)=(\rho/\mu)h_a(n)$. Opposite arrivals have predictable intensity
$\mu$, even though the server's side is adaptive: both independent
streams have that rate, and the current side is determined from their
past. Hence
\[
 \rho n+\mu[V(n+1)-V(n)]\ge\rho g(a).
\]
Integrate over $[0,T]$ and take expectations. The expected sum of
potential increments at opposite arrivals equals the integral of their
predictable intensity times the increment. This compensator identity
follows first for past-measurable step functions from independent
Poisson increments, then for the bounded predictable increment above
by approximation. Boundedness of $V$ and the finite expected number of
arrivals justify the expectation.

Every crossing clears the opposite queue, lowers its potential to zero,
and costs $w\ge V(n)$. Starting from $V(0)=0$, telescoping the arrival
increments and crossing decreases therefore gives
\[
 \E[\text{movement and waiting through }T]+\E[V(n_T)]
 \ge \rho g(a)T.
\]
If $n_T>0$, completion requires another crossing costing at least
$w\ge V(n_T)$; if $n_T=0$ the terminal potential is zero.
All remaining waiting is nonnegative. Thus the full completed cost
satisfies $\E[\ALG^{\rm threshold}_T]\ge\rho g(a)T$.
Lemma~\ref{lem:r5-two-pathwise} then yields
$4\,\E[\OPT^{\rm two}_T]\ge\,\E[\ALG^{\rm threshold}_T]$.
\end{proof}

\subsection{A fixed-home algorithm and its expected cost}\label{os:sec:algorithm}
The two-site analysis leaves two choices to make on a tree: where
should the home be, and how frequently should each vertex be visited?
We first choose the home so that no branch can contain most of the
demand. We then extend the purchase/waiting balance along each root
path, choosing periods that allow nearby visits to share their travel.

\paragraph{Choosing the home.}
With no positive-rate sites, do nothing. With exactly one, move there
on its first request and stay; this is optimal on every finite input.
Henceforth assume at least two distinct vertices have positive rates.
Choosing the vertex of
highest individual rate need not work. Many lower-rate vertices may
form a distant cluster whose combined demand is much larger; returning
from that cluster to the highest-rate vertex would be unnecessarily
expensive. Section~\ref{os:sec:closing} gives a quantitative example.

Instead, retain the minimal subtree spanning the positive-rate vertices,
with physical weight $W$, and let $L=\sum_v\lambda_v$. Choose a
\emph{rate-weighted centroid} $h$: every component left after deleting
$h$ has rate at most $L/2$. Thus each branch has at least as much
arrival rate outside it as inside, as needed for the two-stream lower
bound. Such a vertex exists: whenever one component has more than half
the rate, move into it; the reverse direction then has less than half,
so this walk cannot return along an edge and must terminate.

\paragraph{Choosing the periods.}
Root the retained subtree at $h$, let $w_e=2\ell_e$ be the round-trip
cost of edge $e$, and let $\Lambda_v$ be the rate below vertex $v$.
We use a continuous service scale $q_v$ before rounding it to a period.
Suppose the parent's scale $q_u$ is already chosen. Across edge $uv$,
the sparse-demand choice increases this scale by at most $w_{uv}$.
For frequent demand, we instead balance the edge cost with the increase
in the quadratic waiting expression: the new scale $q_v$ solves
$\Lambda_v(q_v^2-q_u^2)/2=w_{uv}$. Taking the smaller of these two
choices gives
\begin{equation}\label{os:eq:clocks}
 q_h=0,\qquad
 q_v=\min\left\{q_u+w_{uv},\sqrt{q_u^2+\frac{2w_{uv}}{\Lambda_v}}\right\},
 \qquad p_v=2^{\theta+\lceil\log_2(q_v/(2\ln2))-\theta\rceil},
\end{equation}
where $u$ is the parent of $v$ and the common phase $\theta$ is selected below. Zero-rate outer branches have been
removed, so $\Lambda_v>0$. Lemma~\ref{os:lem:clock-balance} will show
how these local balances add up to the cost of the tours.

The values $q_v$ increase along root paths. Rounding the scaled values
$q_v/(2\ln2)$ with a common phase makes each descendant period an integer multiple of
every nonroot ancestor period. We use a common time unit and origin, so a
descendant's tick is also a tick of its nonroot ancestors. At a tick we take
the union of paths to due vertices with pending requests and make
one closed tour of that subtree. Shared edges are paid only once in
each direction, and any other pending requests encountered are served
as well.

\begin{algorithm}[!htbp]
\caption{Centroid-home synchronized clocks for additive online service}
\label{os:alg:osd-tree}
\begin{algorithmic}[1]
\State Prune zero-rate branches; choose a rate-weighted centroid $h$.
\State Root at $h$ and compute subtree rates $\Lambda_v$.
\State Compute $q_v$, choose the common phase as below, and set $p_v$ by~\eqref{os:eq:clocks}.
\State At time $0$, move to $h$; all clocks have time origin $0$.
\Statex \textbf{On an arrival at the server's vertex:}
  \State Serve it immediately.
\Statex \textbf{When one or more clocks ring at time $t$:}
  \State Let $D_t$ be the due vertices with pending requests.
  \If{$D_t\ne\varnothing$}
    \State Let $T_t$ be the union of paths from $h$ to vertices in $D_t$.
    \State Traverse $T_t$ in a closed depth-first tour from $h$.
    \State Serve every pending request encountered, whether its clock is due or not.
  \EndIf
\end{algorithmic}
\end{algorithm}

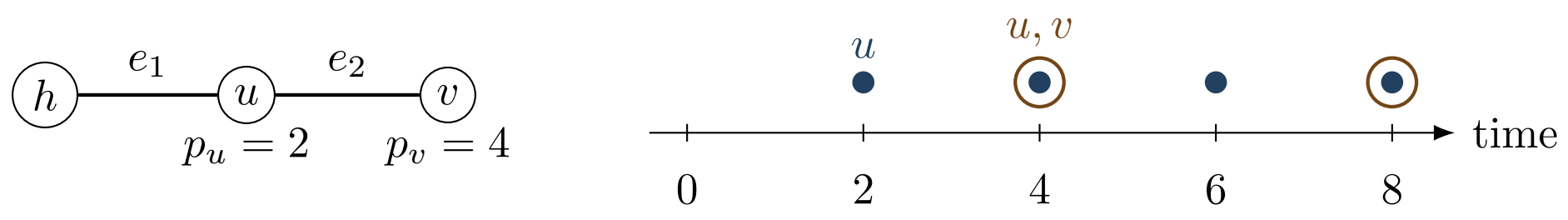
\begin{figure}[t]
\centering
\begin{tikzpicture}[>=Latex,font=\small]
\node[circle,draw,inner sep=2pt] (h) at (0,1.0) {$h$};
\node[circle,draw,inner sep=2pt] (u) at (1.6,1.0) {$u$};
\node[circle,draw,inner sep=2pt] (v) at (3.2,1.0) {$v$};
\draw[thick] (h)--(u) node[midway,above] {$e_1$};
\draw[thick] (u)--(v) node[midway,above] {$e_2$};
\node[below=.13cm] at (u) {$p_u=2$};
\node[below=.13cm] at (v) {$p_v=4$};
\draw[->] (4.8,.7)--(11.2,.7) node[right] {time};
\foreach \x/\lab in {5.1/0,6.5/2,7.9/4,9.3/6,10.7/8}
  {\draw (\x,.63)--(\x,.77);\node[below=.12cm] at (\x,.62) {$\lab$};}
\foreach \x in {6.5,7.9,9.3,10.7}
  {\fill[PaperBlue!70!black] (\x,1.1) circle (2.5pt);}
\foreach \x in {7.9,10.7}
  {\draw[PaperAmber!70!black,thick] (\x,1.1) circle (5.5pt);}
\node[PaperBlue!70!black,above] at (6.5,1.16) {$u$};
\node[PaperAmber!70!black,above] at (7.9,1.25) {$u,v$};
\node[align=center] at (8.0,-.55)
  {At time $4$, one tour to $v$ clears both due locations;\\
   the common edge $e_1$ is traversed only once in each direction.};
\end{tikzpicture}
\caption{Dyadic synchronization on a two-edge rooted path. A tick of
the descendant period $p_v=4$ is also a tick of the ancestor period
$p_u=2$. Empty queues may eliminate a tour, but cannot create an
additional traversal.}
\label{os:fig:clock-sync}
\end{figure}

Figure~\ref{os:fig:clock-sync} illustrates the edge-sharing purpose of
the common time origin. A due request at a descendant can make the
tour traverse its whole root path, and a due request at an ancestor
is cleared on that same tour. The algorithm computes the union of
paths before moving, so it never pays separately for their overlap.

Every request is served by its next own tick, or earlier on another
tour. Finite periods therefore complete every finite input without
an end signal. The initial move costs at most $d(s_0,h)$, already
included in the bound below.

Write $B=\sum_{v\ne h}\lambda_vq_v$. We choose the common phase
using the rates and lengths before execution. Its analysis also
accounts for every final clearing.
\begin{lemma}[Clock balance]\label{os:lem:clock-balance}
For an edge $e=(u,v)$ oriented away from $h$, write
$x=q_u$, $y=q_v$, $\Delta=y-x$, $w=w_e$, and $\Lambda=\Lambda_v$.
Then $0<\Delta\le w$ and
\begin{equation}\label{os:eq:density}
 \frac w\Delta=\max\{1,\Lambda(x+y)/2\},\qquad
 w\min\{\Lambda,1/y\}\le\Lambda\Delta.
\end{equation}
Moreover,
\begin{equation}\label{os:eq:clock-telescope}
 \sum_{v\ne h}\Lambda_v(q_v-q_{\operatorname{par}(v)})
 =\sum_{v\ne h}\lambda_vq_v=B.
\end{equation}
\end{lemma}
\begin{proof}
Both candidates exceed $x$. If $y=x+w$, comparison with the square
root gives $\Lambda(x+y)/2\le1$, and
$w\min\{\Lambda,1/y\}\le w\Lambda=\Lambda\Delta$.
Otherwise $2w=\Lambda(y^2-x^2)$ and
$\Lambda(x+y)/2\ge1$, while
$w\min\{\Lambda,1/y\}\le w/y\le\Lambda\Delta$.
These are the two assertions in~\eqref{os:eq:density}.
Finally, expanding $\Lambda_v$ and exchanging finite sums makes each
type $z$ contribute its rate times the telescoping clock increments
on $h$--$z$, namely $\lambda_zq_z$.
\end{proof}

\subsubsection{Choosing the common clock scale}
Use the periods in~\eqref{os:eq:clocks} and put $a_0=1/(2\ln2)$.
Write
$$
 U(\theta)=\sum_{v\ne h}w_{e_v}\min\{\Lambda_v,1/p_v(\theta)\}
          +\frac12\sum_{v\ne h}\lambda_vp_v(\theta).
$$
For every fixed phase,
$$
 \mathbb E[\operatorname{ALG}_T]
 \le U(\theta)T+K_\theta+d(s_0,h),\qquad
 K_\theta=2\sum_{v\ne h:\lambda_v>0}d(h,v)
              +\frac18\sum_{v\ne h}\lambda_vp_v(\theta)^2.
$$
Here the expectation concerns arrivals only.
To verify the completion term directly, let $n_v$ be the number of
positive-rate descendants of $v$. Every tour crossing $e_v$ clears a
distinct pending arrival below $v$, and every crossing before $T$
occurs at a multiple of $p_v$. After $T$, each positive-rate descendant
can trigger at most one further tour. Thus the number of these tours
is at most both the number of arrivals below $v$ and
$\lfloor T/p_v\rfloor+n_v$.
Taking expectations gives movement at most
$T\sum_v w_{e_v}\min\{\Lambda_v,1/p_v\}+\sum_vw_{e_v}n_v$.
The last sum is $2\sum_{z\ne h:\lambda_z>0}d(h,z)$.
Each arrival is served by its next own tick, and direct integration
of the residual time to that tick gives waiting at most
$\sum_v\lambda_v(Tp_v/2+p_v^2/8)$.
The initial relocation adds at most $d(s_0,h)$ and takes no time. This proves the displayed estimate
with all post-horizon clearings included.
If $\theta$ is temporarily uniform in $[0,1)$, then
$p_v/(a_0q_v)$ is distributed as $2^U$ for uniform $U\in[0,1)$.
Direct integration gives
$$
 \mathbb E_\theta[1/p_v]=\frac1{2a_0q_v\ln2}=\frac1{q_v},\qquad
 \mathbb E_\theta[p_v]=\frac{a_0q_v}{\ln2}.
$$
Since $z\mapsto\min\{\Lambda_v,z\}$ is concave,
Lemma~\ref{os:lem:clock-balance} gives
$$
 \begin{aligned}
 \mathbb E_\theta[U(\theta)]
 &\le\sum_{v\ne h}w_{e_v}\min\{\Lambda_v,1/q_v\}
      +\frac{a_0}{2\ln2}\sum_{v\ne h}\lambda_vq_v\\
 &\le C_0 B,\qquad C_0=1+\frac1{4(\ln2)^2}.
 \end{aligned}
$$
Choose once, from the known rates and lengths, any phase satisfying
$U(\theta)\le C_0B$.
Such a phase exists by the averaging inequality.
This is a finite-dimensional deterministic choice: the finitely many
period-rounding breakpoints, together with the finitely many crossings
$p_v=1/\Lambda_v$, split the phase interval into pieces on which
$U(\theta)=A2^{-\theta}+D2^\theta+E$.
Solving the scalar inequality on these pieces, with the actual endpoint
values of the ceiling function, supplies a valid phase.
No randomization at runtime or information about future arrivals is
needed.

\begin{lemma}[Online finite-horizon bound]\label{os:lem:upper}
The selected phase and Algorithm~\ref{os:alg:osd-tree} satisfy
$$
 \E[\ALG_T]\le C_0BT+K_\theta+d(s_0,h),
 \qquad C_0=1+\frac1{4(\ln2)^2}.
$$
The constant $K_\theta$ is the finite completion term defined above.
\end{lemma}
\begin{proof}
Substitute the selected phase's inequality $U(\theta)\le C_0B$
into its complete movement and waiting estimate.
\end{proof}
\subsection{The clock geometry and analytical marks}

Keep the rooted tree and clocks from~\eqref{os:eq:clocks}.
For an edge with endpoint clocks $x=q_u<y=q_v$, write
$$
 D_e=\frac{w_e}{y-x}
 =\max\{1,\Lambda_e(x+y)/2\},\qquad
 \nu_e=\sum_{z\text{ below }e}\frac{\lambda_z}{q_z}.
$$
These formulas use only positive-rate nonroot vertices in the sum.
The first identity is Lemma~\ref{os:lem:clock-balance}.
Since $q_z\ge y$ below $e$,
$$
 D_e\ge\max\{1,\nu_e y(x+y)/2\}.
$$
Also the finite-sum telescope gives
$$
 \sum_e\nu_e(y^2-x^2)
 =\sum_{z\ne h}\frac{\lambda_z}{q_z}
      \sum_{e\in h\text{--}z}(q_{\rm child}^2-q_{\rm parent}^2)
 =B.
$$

Give each actual request at most one auxiliary target as follows.
For completeness, partition a circle of circumference
$L=\sum_v\lambda_v$ into vertex intervals of lengths $\lambda_v$,
keeping each root branch contiguous.
An occurrence chooses a uniform point in its source interval, shifts
it by $L/2$, and takes the containing vertex as its auxiliary target.
Discard target $h$.
Every branch has rate at most $L/2$, so a nonroot target lies outside
the source's branch.
The target stream at $z$ has rate $\lambda_z$.
Each actual occurrence has one native location and at most one target.
Native requests in any one branch and requests with targets in that
branch use disjoint source processes and are independent.
These labels are used only in the proof and are independent of any
offline schedule chosen from the original input.

\begin{figure}[t]
\centering
\begin{tikzpicture}[>=Latex,font=\small,
 vertex/.style={circle,draw=PaperInk,fill=white,inner sep=2.7pt},
 every path/.style={line cap=round}]
\fill[PaperBlue!7,rounded corners] (-.8,-1.85) rectangle (1.25,1.75);
\fill[PaperTeal!8,rounded corners] (2.0,-1.85) rectangle (4.25,1.75);
\node[vertex,fill=PaperAmber!15] (h) at (0,0) {$h$};
\node[vertex] (a) at (-.15,1.1) {$a$};
\node[vertex] (b) at (.05,-1.1) {$b$};
\node[vertex] (v) at (2.6,0) {$v$};
\node[vertex] (z) at (3.65,1.05) {$z$};
\node[vertex] (r) at (3.65,-1.05) {$r$};
\draw[thick] (a)--(h)--(b);
\draw[very thick,PaperAmber] (h)--(v) node[midway,above] {$e$};
\draw[thick] (z)--(v)--(r);
\draw[dashed,PaperAmber] (1.65,-1.55)--(1.65,1.55);
\node[align=center] at (.1,-2.25) {outside branch\\rate at least $L/2$};
\node[align=center] at (3.1,-2.25) {centroid branch\\rate at most $L/2$};
\draw[->,PaperPurple,thick,bend left=27] (a) to
 node[above,sloped,font=\footnotesize] {auxiliary target} (z);
\draw[->,thick,PaperInk] (4.65,0)--(5.55,0);
\node[vertex,fill=PaperBlue!8] (l) at (6.25,0) {$-$};
\node[vertex,fill=PaperTeal!10] (q) at (9.1,0) {$+$};
\draw[very thick,PaperAmber] (l)--(q)
 node[midway,above] {$\ell_e$};
\node[align=center,text width=2.6cm] at (6.25,-1.1)
 {auxiliary stream\\rate $\mu_e$};
\node[align=center,text width=2.6cm] at (9.1,-1.1)
 {native stream\\rate $\mu_e$};
\node[align=center] at (7.7,1.1) {project the trajectory onto the cut};
\node[align=center] at (7.7,-2.25)
 {each retained request has\\waiting coefficient $\rho_e$};
\end{tikzpicture}
\caption{The centroid supplies demand outside every root branch. A source
outside the branch may receive an auxiliary target $z$ inside it, but
its virtual request remains on the source's side of the cut. Thinning
native and auxiliary occurrences below $e$ gives equally rated,
independent streams; the projected server starts on the side containing
$s_0$. The marks are for analysis only.}
\label{fig:centroid-projection}
\end{figure}
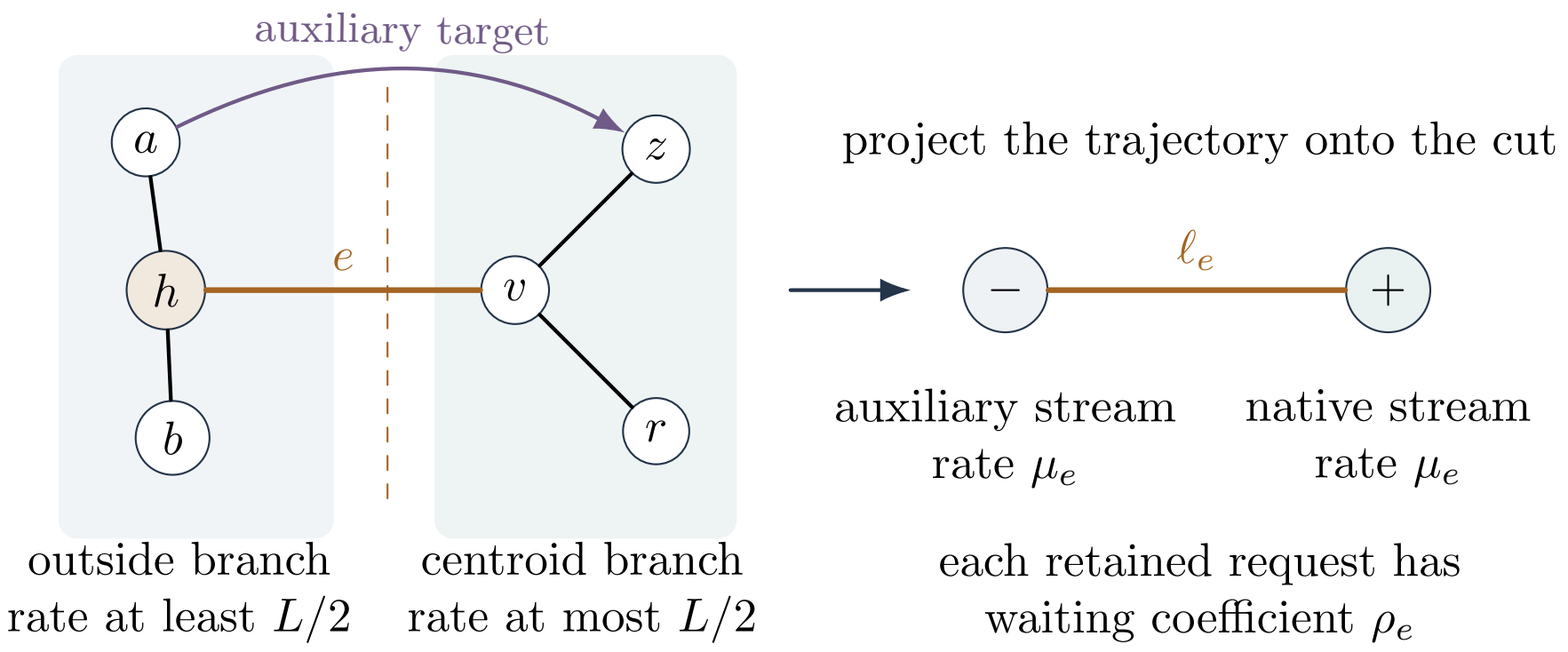

\subsection{One two-site projection per edge}
For an edge $e=(u,v)$, write $x=q_u$, $y=q_v$, and $\Delta=y-x$.
For every native or auxiliary-target occurrence at $z$ below $e$,
retain it for this edge independently with probability $y/q_z$.
This is at most one because $q_z\ge y$.
Create a two-site instance on the sides of $e$ with parameters
$$
 \mu_e=y\nu_e,\qquad \rho_e=\frac{\Delta}{2y},
 \qquad \text{crossing cost }\ell_e=\frac{w_e}{2}.
$$
Native requests lie below the edge, and auxiliary requests lie above
it. Their sources are disjoint, so they form independent Poisson
streams of the same rate $\mu_e$. The two-site server starts on the
side containing the original $s_0$.

\begin{lemma}\label{lem:r5-projection}
For every finite horizon,
$$
 \E[\OPT_T]\ge\sum_e\,\E[\OPT^{\rm two}_{e,T}]
 \ge\frac T8\sum_e\frac{y-x}{y}g(\nu_eD_ey^2).
$$
\end{lemma}
\begin{proof}
Fix the actual input and a completing trajectory before choosing the
analytical marks. Project its position onto the sides of each edge.
A visit to a source serves its virtual request no later than the real
one: native sources are below the edge, and auxiliary sources are
outside its root branch. The initial sides are correct, and the
sum of projected crossing costs is the original movement cost on
the retained tree, hence at most the full movement cost.

Condition on the first-layer targets. A request at $z$ with waiting
$D$ has expected projected waiting in its native role at most
$$
 D\sum_{e\in h\text{--}z}\frac{q_{\rm child}}{q_z}
                 \frac{q_{\rm child}-q_{\rm parent}}{2q_{\rm child}}
 =\frac D{2q_z}\sum_{e\in h\text{--}z}
                 (q_{\rm child}-q_{\rm parent})=\frac D2.
$$
Its one auxiliary role costs at most another $D/2$ by the same
calculation. A root request has no native role. Thus the expected
sum of projected schedule costs is at most the actual cost, including
all completion after $T$. Each virtual optimum is no greater than
its projected schedule cost. Average the marks, take the infimum
over original schedules, and then average the input. This proves
the first inequality without allowing the comparator to observe
the analytical marks.

For the second inequality apply Lemma~\ref{lem:r5-two-lower}, since
$\mu_e\ell_e/\rho_e=\nu_eD_ey^2$ and
$\rho_e/4=(y-x)/(8y)$.
\end{proof}

The remaining comparison needs only AM--GM. For $t\ge0$,
$$
 g(t)\ge\min\{t,\sqrt t\}.
$$
Indeed the $n=1$ term in $g$ is $t$, while for $n\ge2$,
$t/n+(n-1)/2\ge\sqrt{2t(n-1)/n}\ge\sqrt t$.

\begin{theorem}\label{thm:r5-tree-certificate}
For the continuous tree clocks and every $T\ge0$,
$$
 \E[\OPT_T]\ge\frac1{16}BT.
$$
\end{theorem}
\begin{proof}
On an edge put $a=\nu_e y(x+y)/2$ and $t=\nu_eD_ey^2$.
Since $x\le y$ and $D_e\ge\max\{1,a\}$, one has
$\nu_ey^2\ge a$ and therefore $t\ge\max\{a,a^2\}$.
The preceding bound gives $g(t)\ge a$. Substitution in
Lemma~\ref{lem:r5-projection} yields
$$
 \E[\OPT_T]\ge\frac T{16}\sum_e\nu_e(y^2-x^2)
 =\frac T{16}B.
$$
The last equality is the clock telescope established above.
\end{proof}

\begin{theorem}[Mobile service on trees]\label{os:thm:tree}
On every finite tree with at least two positive-rate sites, the
centroid-home policy satisfies
$$
 \E[\ALG_T]\le16C_0\,\E[\OPT_T]
                  +K_\theta+d(s_0,h).
$$
When at least two distinct sites have positive rate, its long-run
ratio is at most $16C_0<24.326$.
\end{theorem}
\begin{proof}
Combine Lemma~\ref{os:lem:upper} with
Theorem~\ref{thm:r5-tree-certificate}. In the nontrivial case $B>0$,
so the offline expectation grows linearly and the fixed completion
term vanishes in the ratio. The algorithm description handles empty
and single-site demand separately.
\end{proof}

\subsection{Why the choice of home matters}\label{os:sec:closing}
For two or more positive-rate sites, Algorithm~\ref{os:alg:osd-tree}
uses a fixed home after its initial relocation. Its guarantees compare
with the unrestricted mobile optimum; they do not identify an
optimal causal rate or an optimal home.

\paragraph{A failed home-selection rule.}
Highest individual rate is insufficient when the number of points grows.
For $m\ge2$, put a site $A$ of rate $1$ at zero, and $m$ sites of rate $1/2$ in an interval
of width $\varepsilon=m^{-4}$ starting at $L=m^2$. Consider policies
consisting of instantaneous closed excursions from $A$, with the server
at $A$ between excursions. Every excursion serving the cluster costs
at least $2L$. 

With cluster rate $\mu=m/2$, use
blocks of half-length $q=\sqrt{L/\mu}$. The one-sided certificate

$$
 \min(2L,qN),
$$
 $N\sim\operatorname{Pois}(\mu q)$, charges either an excursion in the
block or the wait of its first-half arrivals when no excursion occurs.
Different blocks use disjoint movements and request sets.
For $X=qN$, one has $\E[X]=L$ and $\E[X^2]\le2L^2$. Cauchy--Schwarz gives
$$
 \E[\min\{2L,X\}]\ge
 \frac{(\E[X])^2}{\E[X]+\E[X^2]/(2L)}\ge L/2\ge L/3.
$$
Thus any such home policy has rate at least 
$$
 \sqrt{L\mu}/6=\Omega(m^{3/2}).
$$

A home at the near end of the cluster serves its interior by immediate
round trips at rate at most $m\varepsilon$, and visits $A$ with period
$2\sqrt L$, costing rate at most $2\sqrt L=2m$.

The unique highest-rate home therefore loses $\Omega(\sqrt m)$ even against
a causal comparator. A rate median or centroid avoids this example.

The example shows why the home should reflect the distribution of demand
across the tree, rather than the rate at one vertex. Our centroid choice
provides the outside streams needed by the lower bound. It remains
interesting to improve the constants or extend the constant guarantee
to general metrics without a tree-embedding loss. Another direction is
to study how the policy can learn arrival rates from observed requests.

\section{Conclusion}\label{sec:conclusion}
Known Poisson arrival rates allow fixed spatial choices to compete
with an offline optimum that adapts to the realized demand. Demand
radii and a space--time certificate justify a fixed facility assignment.
Connected regions justify fixed roots on a tree, after which the
local rooted-forest lemma supplies the temporal scheduler and the
certificate used to select an assignment. A
rate-weighted centroid and balanced cut projections justify returning
a mobile server to one home, even though the optimum need never return.
These are three complementary ways to control the cost of a spatial
restriction, with complete finite-horizon comparisons in each case.

The multi-root construction also admits polynomial deterministic
preprocessing through a local-marginal LP and a constructive support
decomposition. The facility and mobile algorithms use finite real-arithmetic
constructions; their exact algebraic computations carry no polynomial
bit-complexity claim here. Improving the constants, making the root
preprocessing practical, and extending the tree arguments to general
metrics remain natural questions. Learning arrival rates from observations
would require a further analysis: the present policies know the rates
when they choose their assignments, home, and clocks.

\section*{Declaration of AI use}
Generative AI tools (OpenAI's Astra and Codex) assisted mathematical
exploration, proof development, drafting and revision, literature searches,
and LaTeX preparation. The authors are responsible for the correctness of
the results and references and for the final text.

\begingroup
\small
\bibliographystyle{plainurl}
\bibliography{references}

@inproceedings{bienkowski-et-al-2016,
  author = {Marcin Bienkowski and Martin B\"ohm and Jaros\l{}aw Byrka and Marek Chrobak and Christoph D\"urr and Luk\'a\v{s} Folwarczn\'y and \L{}ukasz Je\.{z} and Ji\v{r}\'{i} Sgall and Nguyen Kim Thang and Pavel Vesel\'y},
  title = {{Online algorithms for multi-level aggregation}},
  booktitle = {24th Annual European Symposium on Algorithms (ESA 2016)},
  pages = {12:1--12:17},
  series = {Leibniz International Proceedings in Informatics (LIPIcs)},
  year = {2016},
  volume = {57},
  publisher = {Schloss Dagstuhl -- Leibniz-Zentrum f{\"u}r Informatik},
  doi = {10.4230/LIPIcs.ESA.2016.12},
}

@inproceedings{buchbinder-2017,
  author = {Niv Buchbinder and Moran Feldman and Joseph (Seffi) Naor and Ohad Talmon},
  title = {{$O(\mathrm{depth})$-competitive algorithm for online multi-level aggregation}},
  booktitle = {Proceedings of the 28th Annual ACM-SIAM Symposium on Discrete Algorithms (SODA)},
  pages = {1235--1244},
  year = {2017},
  doi = {10.1137/1.9781611974782.80},
  publisher = {Society for Industrial and Applied Mathematics},
}

@inproceedings{azar-touitou-2019,
  title = {{General framework for metric optimization problems with delay or with deadlines}},
  author = {Yossi Azar and Noam Touitou},
  booktitle = {2019 IEEE 60th Annual Symposium on Foundations of Computer Science (FOCS)},
  pages = {60--71},
  year = {2019},
  doi = {10.1109/FOCS.2019.00013},
  eprint = {1904.07131},
  archiveprefix = {arXiv},
  publisher = {IEEE},
}

@inproceedings{azar-touitou-2020,
  title = {{Beyond tree embeddings---a deterministic framework for network design with deadlines or delay}},
  author = {Yossi Azar and Noam Touitou},
  booktitle = {2020 IEEE 61st Annual Symposium on Foundations of Computer Science (FOCS)},
  pages = {1368--1379},
  year = {2020},
  eprint = {2004.07946},
  archiveprefix = {arXiv},
  doi = {10.1109/FOCS46700.2020.00129},
  publisher = {IEEE},
}

@inproceedings{bienkowski-et-al-2022,
  author = {Marcin Bienkowski and Martin B\"ohm and Jaros\l{}aw Byrka and Jan Marcinkowski},
  title = {{Online facility location with linear delay}},
  booktitle = {Approximation, Randomization, and Combinatorial Optimization. Algorithms and Techniques (APPROX/RANDOM 2022)},
  pages = {45:1--45:17},
  series = {Leibniz International Proceedings in Informatics (LIPIcs)},
  year = {2022},
  volume = {245},
  publisher = {Schloss Dagstuhl -- Leibniz-Zentrum f{\"u}r Informatik},
  doi = {10.4230/LIPIcs.APPROX/RANDOM.2022.45},
}

@inproceedings{mari-et-al-2023,
  author = {Mathieu Mari and Micha\l{} Paw\l{}owski and Runtian Ren and Piotr Sankowski},
  title = {{Online matching with delays and stochastic arrival times}},
  booktitle = {Proceedings of the 22nd International Conference on Autonomous Agents and Multiagent Systems (AAMAS)},
  pages = {976--984},
  year = {2023},
  url = {https://www.ifaamas.org/Proceedings/aamas2023/pdfs/p976.pdf},
  eprint = {2210.07018},
  archiveprefix = {arXiv},
  publisher = {International Foundation for Autonomous Agents and Multiagent Systems},
}

@inproceedings{mari-et-al-2024,
  author = {Mathieu Mari and Micha\l{} Paw\l{}owski and Runtian Ren and Piotr Sankowski},
  title = {{Online multi-level aggregation with delays and stochastic arrivals}},
  booktitle = {35th International Symposium on Algorithms and Computation (ISAAC 2024)},
  pages = {49:1--49:20},
  series = {Leibniz International Proceedings in Informatics (LIPIcs)},
  year = {2024},
  volume = {322},
  publisher = {Schloss Dagstuhl -- Leibniz-Zentrum f{\"u}r Informatik},
  doi = {10.4230/LIPIcs.ISAAC.2024.49},
}

@article{connected,
  author = {Matthias Englert and Anupam Gupta and Robert Krauthgamer and Harald R\"acke and Inbal Talgam-Cohen and Kunal Talwar},
  title = {{Vertex sparsifiers: New results from old techniques}},
  year = {2014},
  url = {https://www.inbaltalgam.com/papers/Sparsifiers.pdf},
  journal = {SIAM Journal on Computing},
  volume = {43},
  number = {4},
  pages = {1239--1262},
  doi = {10.1137/130908440},
}

@inproceedings{azar-ganesh-ge-panigrahi-2017,
  author = {Yossi Azar and Arun Ganesh and Rong Ge and Debmalya Panigrahi},
  title = {{Online service with delay}},
  booktitle = {Proceedings of the 49th Annual ACM SIGACT Symposium on Theory of Computing (STOC)},
  pages = {551--563},
  year = {2017},
  doi = {10.1145/3055399.3055475},
  eprint = {1708.05611},
  archiveprefix = {arXiv},
  publisher = {ACM},
}

@inproceedings{touitou-2023,
  title = {{Improved and deterministic online service with deadlines or delay}},
  author = {Noam Touitou},
  booktitle = {Proceedings of the 55th Annual ACM Symposium on Theory of Computing (STOC)},
  pages = {761--774},
  year = {2023},
  eprint = {2306.05744},
  archiveprefix = {arXiv},
  doi = {10.1145/3564246.3585107},
  publisher = {ACM},
}

@inproceedings{touitou-service-2025,
  author = {Noam Touitou},
  title = {{Nearly-optimal algorithm for non-clairvoyant service with delay}},
  booktitle = {42nd International Symposium on Theoretical Aspects of Computer Science (STACS 2025)},
  pages = {74:1--74:21},
  series = {Leibniz International Proceedings in Informatics (LIPIcs)},
  year = {2025},
  volume = {327},
  publisher = {Schloss Dagstuhl -- Leibniz-Zentrum f{\"u}r Informatik},
  doi = {10.4230/LIPIcs.STACS.2025.74},
}
\endgroup

\appendix
\section{Connected terminal assignment}\label{app:partition}\label{sec:r4-projection}
We specialize the growing connected $0$-extension process of Englert
et al.~\cite[Section~4, Algorithm~1]{connected} to the augmented tree.
The growing assignment construction is due to their work. Here we use
randomly shifted depth partitions, a finite mixture of quadrupling
scales, and a joint endpoint analysis to derive the explicit
service-projection constant $313/8$. We give the tree implementation
and its analysis in full.

Let $T$ be a finite tree with positive edge lengths and terminal set
$K\ne\varnothing$. The assignment is chosen independently of requests,
and the service bound holds for every fixed service. Zero-length edges
are contracted as in the multi-root model.
\subsection{The partition and its elementary properties}

If no positive edge remains after contraction, the assertion is trivial.
Let $m>0$ be the shortest positive edge length.
Fix a depth origin. At scale $\Delta$, choose an independent uniform
offset in $[0,\Delta/2)$ and delete edges meeting an offset depth
level modulo $\Delta/2$. Components are connected, have diameter at
most $\Delta$, and separate a fixed pair $u,v$ with probability at most
$$
 \min\{2d_T(u,v)/\Delta,1\}.
$$
Indeed an edge is cut with probability at most its length divided
by $\Delta/2$, and a union bound applies along the unique path.
The diameter bound follows because a component lies in one depth
slab of width $\Delta/2$ and every path between its vertices stays
in the component. Once $\Delta$ is at least the tree diameter, use
the single whole-tree cluster instead.

Initially assign each terminal its own label. Choose $J$ uniformly from $\{0,\ldots,63\}$ and use scales $\Delta_i=(m/2)4^{i+J/64}$ for $i=0,1,\ldots$, with a fresh independent partition at each round.
Freeze the previously assigned vertices. In each cluster touching
an assigned vertex, remove all frozen vertices, and assign each
remaining connected component the label of one frozen neighbor,
using a fixed tie rule. Other vertices remain unassigned.
Assigned labels never change. The whole-tree round terminates
the process.

Every fiber remains connected because a new component is attached
to an existing vertex with the same label. With $\rho=4/3$, a vertex
first assigned in round $i$ has
$$
 d_T(v,f(v))\le\sum_{h=0}^i\Delta_h<\rho\Delta_i.
$$
This follows by induction through the assigned neighbor in its cluster.
If one endpoint of an uncut edge was assigned before a round, its
other endpoint is assigned by the end of that round. If both endpoints
were unassigned and the edge is uncut, they belong to the same remaining
component, so any labels newly given to them are equal.

The offsets have a finite exact representation: sort vertex depth residues
modulo $\Delta/2$ and sample an open cell with probability proportional
to its length. Its midpoint gives the identical edge partition.
There are finitely many rounds and cells; no continuous sampling
or limiting partition is required. The phase sampler below explains exact sampling of the finitely many algebraic cell probabilities.

\subsection{The cost of waiting for the second endpoint}

Fix an edge of length $\ell$. If exactly one endpoint is assigned,
the other remains unassigned after a later round only if that edge
is cut in that round. Conditional on the previous history, this has
probability at most
$$
 p_j=\min\{1,2\ell/\Delta_j\}.
$$
Independent fresh offsets give products of these bounds across rounds.
Define the two convergent positive series
$$
 S_4=\sum_{k=0}^\infty4^{-k(k+1)/2},\qquad
 B_4=2+\sum_{n=1}^\infty(8/3)^n4^{-n(n+1)/2},
 \qquad H_4=2(4+3S_4).
$$
Elementary geometric tail estimates give
$$
 S_4\le1+\frac14+\frac1{64}+\frac1{4032}<1.266,
 \qquad
 B_4\le2+\frac23+\frac19+\frac1{216}+\frac1{20520}<2.783.
$$
For the first tail, successive terms after $1/64$ have ratio at most
$1/64$. For the second, successive terms after $1/216$ have ratio at
most $1/96$. The displayed rational bound also gives $B_4<64/23$. In particular $H_4<15.596<15.6$.

\begin{lemma}\label{lem:r4-second}
Suppose the first endpoint is assigned in round $i\ge0$, and let $J$
be the round when the second endpoint is assigned. Conditional on
any realized history through round $i$,
$$
 \mathbb E[\Delta_J\mid\text{history}]\le4\Delta_i+H_4\ell.
$$
If one endpoint is initially a terminal and the other is not, then
$\mathbb E[\Delta_J]\le H_4\ell$.
\end{lemma}
\begin{proof}
There is nothing to prove if both endpoints were assigned together.
For a round $j$ with $\Delta_j=2\ell x$ and $x\ge1$, suppose an endpoint
is still unassigned immediately before that round. Products of the
cut probabilities and the identity for the expectation of a
geometrically spaced nonnegative variable give
$$
 \mathbb E[\Delta_J]
 \le 2\ell f(x),\qquad
 f(x)=x+3\sum_{k=0}^\infty4^{-k(k-1)/2}x^{-k}.
$$
Explicitly, $\Pr(J>j+k)$ is at most
$x^{-(k+1)}4^{-k(k+1)/2}$. Substituting this into
$\Delta_j+\sum_{k\ge0}3\Delta_j4^k\Pr(J>j+k)$ proves the display.
If the endpoint was already assigned before $j$, the same upper bound
holds by replacing its assignment scale by the larger scale $\Delta_j$.
The whole-tree terminal round can only decrease these bounds.

The function $f$ is convex on $[1,4]$ because it is a sum of a linear
term and positive multiples of nonnegative inverse powers.
The two endpoints have the same value,
$$
 f(1)=f(4)=4+3S_4.
$$
Thus $2\ell f(x)\le H_4\ell$ for $1\le x\le4$.

Let $j\ge i+1$ be the first subsequent round with $\Delta_j\ge2\ell$.
If $j>i+1$, then $1\le x<4$, so the preceding bound applies, after
dominating any earlier settlement by $\Delta_j$.
If $j=i+1$, then $x\ge1$ and
$$
 2\ell f(x)\le\Delta_j+6\ell(1+S_4)
 \le4\Delta_i+H_4\ell.
$$
For an initially terminal endpoint, take the first $j\ge0$ with
$\Delta_j\ge2\ell$. When $j>0$, again $x<4$.
When $j=0$, the initial scale is less than $2m\le2\ell$, so this threshold case does not occur. This proves the terminal assertion as well.
\end{proof}

\subsection{Projection without an extra additive factor}

For a fixed service subtree $S$ containing terminal $r_0$, put
$$
 U=S\cup\!\bigcup_{uv\in E(S)}P_T(f(u),f(v)).
$$
Following the image paths along an original path from $r_0$ shows
that $U$ is connected and contains $x$ and $f(x)$ for every $x\in S$.
It therefore contains each required path $P_T(x,f(x))$.
Paths assigned to different roots are edge-disjoint because the
fibers are connected and hence path-convex in a tree.
Thus the total assigned purchase cost is at most $w(U)$.

If $f(u)\ne f(v)$, connectedness of the two fibers places their roots
on opposite sides of $T-uv$. Hence $P_T(f(u),f(v))$ contains the
original edge $uv$. Its contribution outside $S$ is at most
$d_T(f(u),f(v))-w_{uv}$. Summing these outside contributions gives
the sharper deterministic conversion
$$
 \sum_{r\in K}w\left(\bigcup_{x\in A:f(x)=r}P_T(x,r)\right)
 \le w(S)+\sum_{\substack{uv\in E(S)\\f(u)\ne f(v)}}
       [d_T(f(u),f(v))-w_{uv}]
 =\sum_{e\in E(S)}W_f(e).
$$
\subsection{A finite mixture of scales}\label{sec:r5-phase-projection}

The finite phase mixture gives the following uniform projection bound.

\begin{theorem}\label{thm:r5-projection}
For every finite edge-weighted tree and nonempty terminal set, there
is a finitely supported distribution of connected terminal
retractions such that, for every fixed connected service subtree $S$
containing a terminal and every client set $A\subseteq V(S)$,
$$
 \mathbb E_f\left[\sum_{r\in K}w\left(\bigcup_{x\in A:f(x)=r}P_T(x,r)\right)\right]
 \le\frac{313}{8}w(S).
$$
Consequently, for every fixed finite input,
$$
 \mathbb E_f[\operatorname{OPT}_f(I)]
 \le\frac{313}{8}\operatorname{OPT}_{\rm free}(I).
$$
The assignment is sampled once, independently of the input.
\end{theorem}

\subsubsection{Finite sampling}
Use the sampler defined at the start of this appendix. Its initial
scale is less than $2m$, so every edge is cut in that round unless
the whole-tree override already terminates the process.
For rational edge lengths, scales and offset-cell probabilities are
algebraic because $4^{J/64}=2^{J/32}$. Fair bits sample them exactly:
refine a binary uniform interval and certified cumulative-probability
intervals until one cell is determined. With finitely many boundaries,
the probability of needing more than $n$ bits is $O(2^{-n})$;
boundary equality is a null event. Thus the sampler has finite support
and terminates almost surely, without a continuous phase limit.

\subsubsection{A joint bound for an edge}

Fix an edge $e=v_0v_1$ of length $\ell$ whose endpoints are not terminals, and
let $A>0$ be their smaller distance to the terminal set.
Put $a=A/\ell$.
Conditional on the chosen phase, let $i_*$ be the first scale with
$\rho\Delta_{i_*}\ge A$ and set
$$
 u=\frac{\Delta_{i_*}}{2A}\in[3/8,3/2),\qquad
 t_k=\frac{\Delta_{i_*+k}}{2\ell}=au4^k.
$$
Earlier scales cannot assign either endpoint by the radius bound.
Let $q_k$ be the probability that the first endpoint-assignment round
is $i_*+k$ and that the edge is cut in that round.
The events defining $q_k$ are disjoint, so $\sum_kq_k\le1$.
The untouched-history product and the fresh edge-cut bound give
$$
 0\le q_k\le c_k(a,u):=
 U_k(u)\min\{1,1/t_k\},\qquad
 U_k(u)=\prod_{h=0}^{k-1}\min\{1,(u4^h)^{-1}\},\qquad U_0=1.
$$
These are unconditional probabilities after fixing the phase; no
fresh partition is conditioned on a future label event.

Define $F(t)$ as the following dominating last-assignment scale,
normalized by $2\ell$, when the first endpoint has just been
assigned at a scale $2\ell t$.
In every later round, let settlement occur at the first uncut edge,
using independent cut probabilities $\min\{1,1/(4^h t)\}$.
This process can only settle later than the actual second endpoint.
Its expectation is finite and satisfies
$$
 F(t)=
 \begin{cases}
 F(4t),&4t\le1,\\
 4t-1+F(4t)/(4t),&4t\ge1.
 \end{cases}
$$
If both endpoints are assigned in the first round, the same bound
still applies because $F(t)\ge4t>t$.
For $t\ge1$, the cut-product series gives
$$
 F(t)=4t+3\sum_{n=0}^{\infty}t^{-n}4^{-n(n+1)/2}
 \le4t+\frac3{1-1/(4t)}\le4t+4.
$$

As in Section~\ref{sec:r4-projection}, use the mixed edge cost
$$
 W_f(e)=\ell\mathbf1_{\{f(v_0)=f(v_1)\}}
       +d_T(f(v_0),f(v_1))\mathbf1_{\{f(v_0)\ne f(v_1)\}}.
$$
Different final labels require the first-assignment cut event.
Conditioning only on the history through that first assignment yields
$$
 \frac{\mathbb E[W_f(e)\mid\text{phase}]}\ell
 \le1+\sum_{k\ge0}q_k L_k(a,u),\qquad
 L_k(a,u)=2\rho[t_k+F(t_k)].
$$
The same disjoint masses $q_k$ pay both assignment radii.

\subsubsection{Two analytic parameter ranges}

Retain $H_4<15.6$ and $B_4\le64/23$ from the geometric estimates
in Section~\ref{sec:r4-projection}.
If $a\le1/8$, compare the closer endpoint directly with its fixed
nearest terminal.
It is assigned no later than the first round in which this pair is
not separated.
The same first-noncut calculation, now with distance $A$, gives
$\mathbb E[\Delta_I]\le H_4A$.
The initial scale is below $2m\le2A$, so the first threshold scale
has ratio in $[1,4)$ as required by that calculation.
The conditional second-endpoint bound gives
$\mathbb E[\Delta_J]\le4\,\mathbb E[\Delta_I]+H_4\ell$.
Therefore
$$
 \frac{\mathbb E[W_f(e)]}\ell
 \le1+\rho H_4(1+5a)<34.8<313/8.
$$

If $a\ge32$, then $t_k\ge au\ge12$ for every $k$.
Thus
$$
 F(t_k)\le4t_k+144/47,\qquad
 \sum_k t_kq_k\le\sum_k U_k(u)\le64/23,
 \qquad
 \sum_kq_k\le\sum_k\frac1{au4^k}\le1/9.
$$
It follows that
$$
 \frac{\mathbb E[W_f(e)]}\ell
 \le1+\frac83\left(5\cdot\frac{64}{23}+\frac{144}{47\cdot9}\right)
 =\frac{42169}{1081}<313/8.
$$
These two bounds hold separately for every initial phase.

\subsubsection{The compact-range integer certificate}

It remains to treat $1/8\le a\le32$.
Partition this interval into the 512 closed bins
$$
 a\in\left[\frac18 2^{h/64},\frac18 2^{(h+1)/64}\right],
 \qquad h=0,\ldots,511,
$$
and partition $[3/8,3/2]$ into the 64 phase bins
$$
 u\in\left[\frac38 4^{j/64},\frac38 4^{(j+1)/64}\right],
 \qquad j=0,\ldots,63.
$$
For a fixed edge, the 64 possible initial phases give exactly one
relative value $u$ in each of these half-open phase bins, after a
cyclic relabeling.
This follows by taking logarithms base four: the values are a common
translate of the 64 equally spaced points modulo one.
Explicitly, for $L=\log_4(3A/(2m))$, the logarithm
$\log_4(u/(3/8))$ is $J/64-L$ modulo one.
$A\ge m$ ensures that the corresponding first admissible round index
is nonnegative, so truncating the sampler at round zero causes no exception.
Thus averaging upper bounds for the 64 bins is valid for the actual
finite phase sampler, not merely for a continuous phase distribution.

Here is the complete finite enclosure used in each rectangular bin.
Let $a_-,a_+,u_-,u_+$ be outward bounds for its endpoints, and put
$t_k^- =a_-u_-4^k$, $t_k^+=a_+u_+4^k$.
Since $U_k$ and each cut factor decrease in $a$ and $u$, valid
capacity bounds are
$$
 \overline c_k=
 \left[\prod_{h=0}^{k-1}\min\{1,(u_-4^h)^{-1}\}\right]
 \min\{1,(t_k^-)^{-1}\}.
$$
Keep $k=0,\ldots,15$.
At the last kept scale use the valid terminal upper bound
$\overline F_{15}=4t_{15}^++4$.
Propagate backwards using the recurrence for $F$.
If $x=4t$ lies entirely below one, keep the previous upper bound.
Above one, maximize $x-1+\overline F_{k+1}/x$ at the interval
endpoints; this is valid because the expression is convex in $x>0$.
For a bin crossing one, include the endpoint $x=1$.
Set $\overline L_k=(8/3)(t_k^++\overline F_k)$.
Finally maximize
$$
 1+\sum_{k=0}^{15}z_k\overline L_k
 \quad\text{subject to}\quad
 0\le z_k\le\overline c_k,\qquad\sum_kz_k\le1.
$$
Sorting the 16 prices in decreasing order and filling the largest
prices first solves this finite linear program exactly.
Relaxing the tail's use of probability mass only enlarges the bound.

The omitted tail is rigorously negligible.
For $k\ge16$ and $u\ge3/8$,
$$
 U_k(u)\le(8/3)^{k-1}4^{-k(k-1)/2},
 \qquad U_{16}<2^{-210},
 \qquad U_{k+1}/U_k\le4^{-15}.
$$
Also $t_k\ge1$, so $L_k\le24t_k$ and $q_kL_k\le24U_k$.
Consequently the full omitted contribution is below $2^{-204}$.

A finite exact-arithmetic computation evaluates these enclosures on
the grid $2^{-96}\mathbb Z$. The endpoints involving $2^{1/64}$ are
bracketed by integer 64th-power comparisons.
Lower products are rounded down, and upper products and divisions
are rounded up by integer quotient and remainder.
No floating-point operation is used.
One additional $2^{-96}$ unit covers the omitted tail.
Averaging the 64 phase-bin bounds for each of the 512 distance bins
gives a finite list of rational upper bounds. Their computed maximum
is at most the following explicit rational number:
$$
 \frac{3099078927284629529722094560869}{2^{96}}
 <\frac{313}{8}.
$$
The strict comparison is an integer comparison.
The largest bin has index $h=263$.
This finite verification completes the compact-range bound.
The exact-arithmetic checker and its complete list of rational bounds
are supplied as ancillary files in \texttt{anc/certificates/}.
They reproduce this finite verification; the enclosure and tail
arguments above justify its use in the proof.

\subsubsection{Terminal edges and the service projection}

If one endpoint is a terminal, the first-assignment radius is zero.
The initial scale is below $2m\le2\ell$, so the same threshold
calculation gives $\mathbb E[\Delta_J]\le H_4\ell$ for every phase.
Thus $\mathbb E[W_f(e)]\le(1+\rho H_4)\ell<21.8\ell$.
If both endpoints are terminals, $W_f(e)=\ell$.
All edges therefore satisfy
$\mathbb E[W_f(e)]\le(313/8)\ell$.

The deterministic conversion established above gives
$$
 \sum_{r\in K}w\left(\bigcup_{x\in A:f(x)=r}P_T(x,r)\right)
 \le\sum_{e\in E(S)}W_f(e).
$$
Taking expectations proves the service bound.
For an arbitrary fixed free-root schedule, split each batch by its
assigned terminal and keep all service times unchanged.
Spatial cost increases by at most $313/8$ in expectation, while
additive waiting is unchanged.
Taking the infimum over the original schedule proves
Theorem~\ref{thm:r5-projection}.

\end{document}